\pdfoutput=1

\documentclass[11pt,twoside,a4paper,cmspaper,final,collab]{cms-tdr}
\def\svnVersion{5e2914e-D}\def\svnDate{2020/03/12}\def\cmsCernNoTag{CERN-EP-2019-254}\def\cmsCernDate{\today}\def\cmsMessage{"Published in the Journal of High Energy Physics as \href{http://dx.doi.org/10.1007/JHEP03(2020)055}{\doi{10.1007/JHEP03(2020)055}.}"}

\begin{document}\cmsNoteHeader{HIG-18-012}

\hyphenation{had-ron-i-za-tion}
\hyphenation{cal-or-i-me-ter}
\hyphenation{de-vices}
\RCS$HeadURL: svn+ssh://svn.cern.ch/reps/tdr2/papers/HIG-17-006/trunk/HIG-17-006.tex $
\RCS$Id: HIG-17-006.tex 420741 2017-08-11 15:28:51Z alverson $
\newlength\cmsFigWidth
\ifthenelse{\boolean{cms@external}}{\setlength\cmsFigWidth{0.85\columnwidth}}{\setlength\cmsFigWidth{0.4\textwidth}}
\ifthenelse{\boolean{cms@external}}{\newcommand{\cmsLeft}{upper\xspace}}{\newcommand{\cmsLeft}{left\xspace}}
\ifthenelse{\boolean{cms@external}}{\newcommand{\cmsRight}{lower\xspace}}{\newcommand{\cmsRight}{right\xspace}}
\newlength\cmsTabSkip\setlength{\cmsTabSkip}{2ex}
\providecommand{\cmsTable}[1]{\resizebox{\textwidth}{!}{#1}}

\newcommand{\mll}{\ensuremath{m_{\ell\ell}}\xspace}
\providecommand{\Pj}{{\HepParticle{j}{}{}}\Xspace}
\providecommand{\PV}{{\HepParticle{V}{}{}}\Xspace}
\newcommand{\mA}{\ensuremath{{m}_{\PSA}}\xspace}
\newcommand{\mH}{\ensuremath{{m}_{\PH}}\xspace}
\newcommand{\mjj}{\ensuremath{m_{\Pj\Pj}}\xspace}
\newcommand{\mlljj}{\ensuremath{m_{\ell\ell\Pj\Pj}}\xspace}

\cmsNoteHeader{HIG-18-012}
\title{Search for new neutral Higgs bosons through the $\PH \to \PZ \PSA \to \ell^{+}\ell^{-} \bbbar$ process in $\Pp \Pp$ collisions at $\sqrt{s} = 13\TeV$}

\date{\today}

\abstract{
This paper reports on a search for an extension to the scalar sector of the standard model, where a new CP-even (odd) boson decays to a \PZ boson and a lighter CP-odd (even) boson, and the latter further decays to a {\cPqb} quark pair. The \PZ boson is reconstructed via its decays to electron or muon pairs.
The analysed data were recorded in proton-proton collisions at a center-of-mass energy $\sqrt{s} = 13\TeV$, collected by the CMS experiment at the LHC during 2016, corresponding to an integrated luminosity of 35.9\fbinv. Data and predictions from the standard model are in agreement within the uncertainties. Upper limits at 95\% confidence level are set on the production cross section times branching fraction, with masses of the new bosons up to 1000\GeV. The results are interpreted in the context of the two-Higgs-doublet model.
}

\hypersetup{%
pdfauthor={CMS Collaboration},%
pdftitle={Search for new neutral Higgs bosons through the H -> ZA -> l+l-bbbar process in pp collisions at sqrts = 13 TeV},
pdfsubject={CMS},%
pdfkeywords={CMS, physics, Higgs}}

\maketitle

\section{Introduction}

The CMS and ATLAS experimental programmes are focusing efforts on the measurement
of the properties of the Higgs boson discovered in 2012~\cite{HiggsAtlas, HiggsCms, HiggsCms2},
which has a mass of about 125\GeV~\cite{Aad:2015zhl, Sirunyan:2017exp, Aaboud:2018wps}.
All measurements to date are consistent with the expectations for a standard model (SM) Higgs boson within the experimental uncertainties.

Additional Higgs bosons are predicted in several extensions of the SM. Examples of these extensions are the two-Higgs-doublet model (2HDM)~\cite{Branco:2011iw},
whose phenomenology is based on the presence of an additional scalar Higgs doublet, and the minimal supersymmetric extension of the SM (MSSM)~\cite{Csaki:1996ks}, which is a particular realisation of the 2HDM. The two Higgs doublets entail the presence of five physical states: two neutral and CP-even
bosons (\Ph and \PH); a neutral and CP-odd boson (\PSA); and two
charged scalar bosons ($\PH^{\pm}$).
Under particular theoretical assumptions~\cite{Branco:2011iw}, the model is often described by the following parameters: the mass of the CP-even boson \PH, \mH; the mass of the pseudoscalar \PSA, \mA; the ratio of the vacuum expectation values of the two doublets, $\tan\beta$; the mixing angle $\alpha$ between the two CP-even bosons; and the soft-breaking term, $m_{12}^{2}$.

Different couplings of the two doublets to right-handed quarks and charged leptons are predicted in various formulations of the 2HDM: in the
Type-I formulation, all fermions couple to only one Higgs doublet; in the Type-II formulation, the up-type quarks couple to a different doublet than the down-type quarks and leptons; in the ``lepton-specific" formulation, the quarks couple to one of the Higgs doublets and the leptons couple to the other; and in the ``flipped" formulation, the up-type quarks and leptons couple to one of the Higgs doublets, while the down-type quarks couple to the other.

\begin{figure}[!htb]
  \centering
    \includegraphics[width=0.44\textwidth]{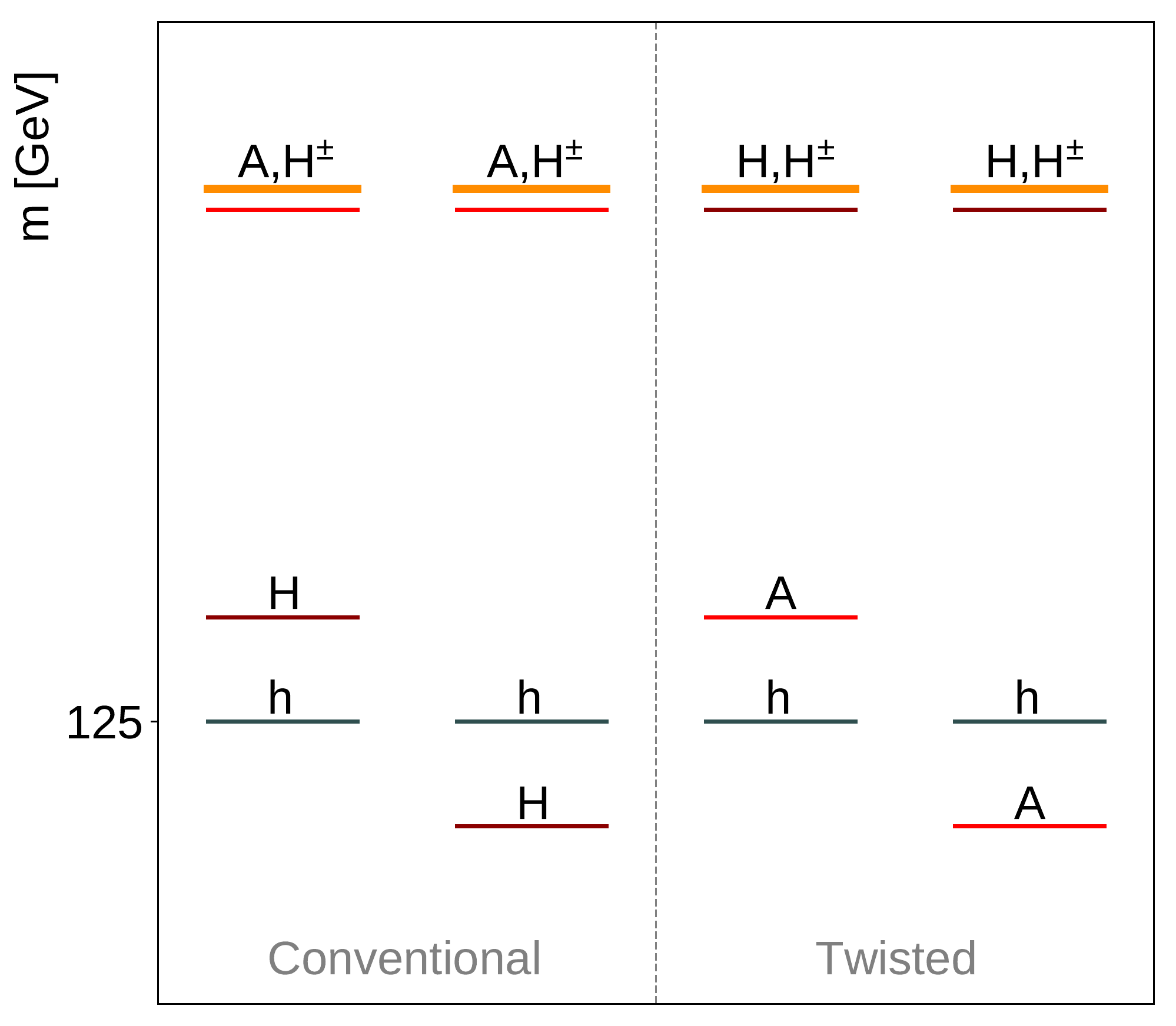}
    \caption{
      Possible 2HDM mass hierarchies: conventional, where \PSA is degenerate in mass with the charged scalars; and twisted~\cite{Gerard:2007kn}, where \PH is degenerate in mass with the charged scalars. In both scenarios, the lighter of the \PSA and \PH bosons can be either heavier or lighter than the observed Higgs boson \Ph{}(125).
     }
    \label{fig:hierarchies_BR}

\end{figure}

\begin{figure}[!htb]
  \centering
    \includegraphics[width=0.98\textwidth]{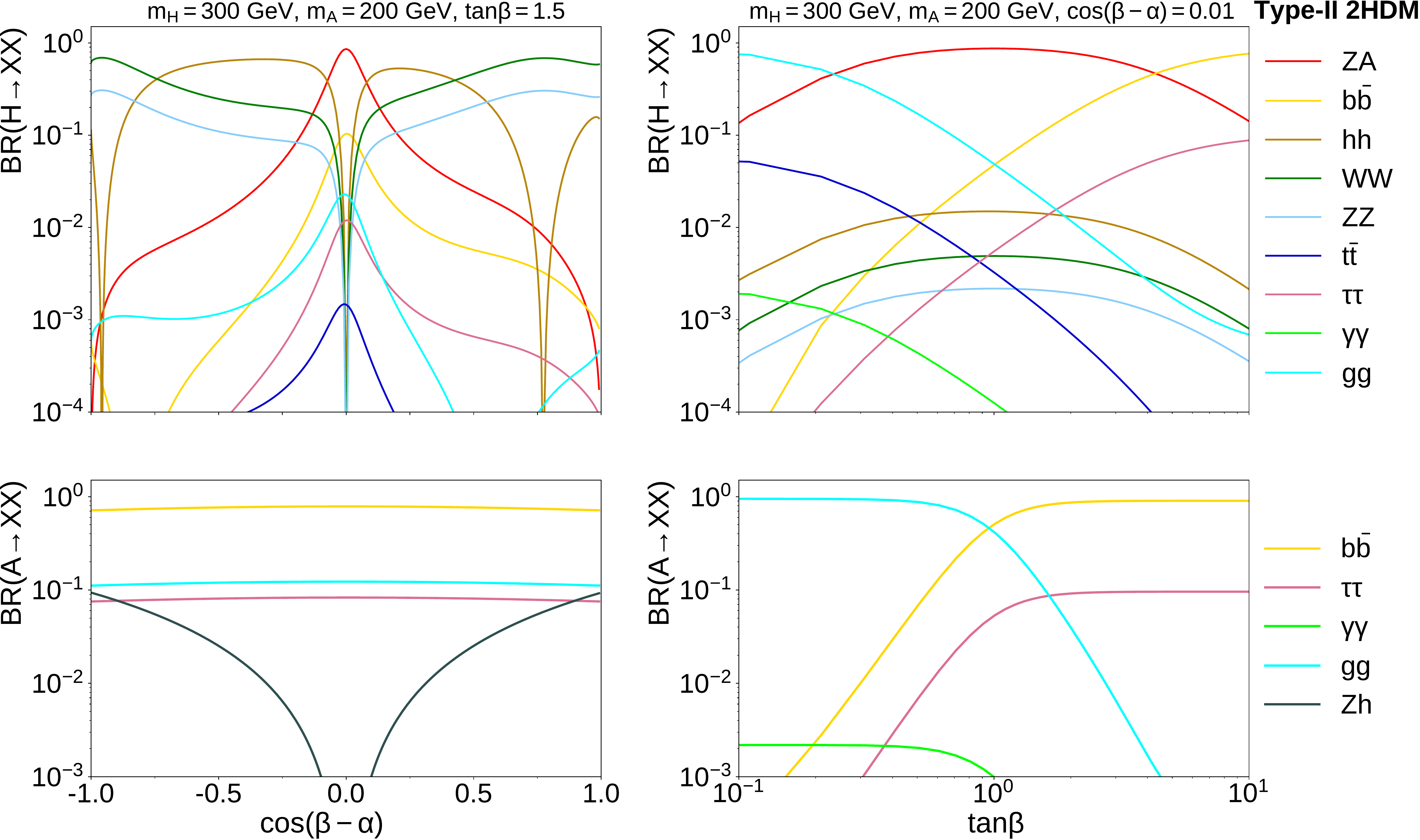}
    \caption{
      The \PH and \PSA branching fractions as a function of $\cos(\beta - \alpha)$ in Type-II 2HDM for
      the following set of parameters: $\tan\beta = 1.5$, $\mH =
      300$\GeV, $\mA = 200$\GeV (left).
      The \PH and \PSA branching fractions as a function of $\tan\beta$ in Type-II 2HDM for the
      following set of parameters: $\cos(\beta - \alpha) = 0.01$, $\mH =
      300$\GeV and $\mA = 200$\GeV (right).
     }
    \label{fig:BR_xsec}

\end{figure}

Different models and assumptions also alter the mass hierarchies, as shown in
Fig.~\ref{fig:hierarchies_BR}. There, and in the rest of the paper, \Ph is identified with the observed Higgs boson. Two scenarios are possible. In the conventional scenario, the pseudoscalar is degenerate in mass with the charged scalars and is heavier than the scalar $\PH$, thus allowing for the $\PSA \to \PZ\PH$ process; while in the twisted~\cite{Gerard:2007kn} scenario, the scalar $\PH$ is degenerate in mass with the charged scalars and is heavier than the pseudoscalar, thus allowing for the $\PH \to \PZ\PSA$ process.
Moreover, in the parameter space region where $\cos (\beta - \alpha)$ approaches 0, the CP-even \Ph has properties indistinguishable from a SM Higgs boson with the same mass. In this
region, known as the alignment limit, the branching fraction of the
heavy scalar \PH to a \PZ boson and a lighter pseudoscalar \PSA is the largest. The branching fractions for several decay channels of the \PH and \PSA bosons for $\mH = 300$\GeV and $\mA = 200$\GeV are shown in Fig.~\ref{fig:BR_xsec} as functions of $\cos (\beta - \alpha)$ (left) and $\tan\beta$ (right).

This paper reports on a search for a new CP-even (odd) neutral Higgs boson decaying
into \PZ and a lighter CP-odd (even) neutral Higgs boson, where the \PZ decays into an opposite-sign electron or muon pair, and the light Higgs boson into a {\cPqb} quark pair. The analysis is performed under the assumption of the twisted mass hierarchy scenario, and subsequently extended to the conventional scenario by interchanging the masses of the two bosons. While this is not crucial when setting model independent upper limits, it appears particularly interesting for theoretical interpretation of the results in the context of the 2HDM, since the theoretical cross sections differ in the two scenarios. The search is based on LHC proton-proton ($\Pp\Pp$) collision data at a center-of-mass energy $\sqrt{s}=13$\TeV collected by the
CMS experiment during 2016, corresponding to an integrated luminosity of
35.9\fbinv. The analysis exploits the invariant mass distributions
of the $\ell\ell\bbbar$, with $\ell$ being an electron or muon, and \bbbar systems to search for a
resonant-like excess of events compatible with the \PH and \PSA masses.

Searches for $\PH \to \PZ\PSA$ production in the same final state have been performed at ${\sqrt{s}=13}$\TeV~\cite{Aaboud:2018eoy} by the ATLAS Collaboration and at ${\sqrt{s}=8}$\TeV~\cite{Khachatryan:2016are} by the CMS Collaboration. The search for $\PSA \to \PZ\Ph$, where \Ph is the observed CP-even boson with mass of about 125\GeV, has been also performed by the ATLAS Collaboration at 8\TeV~\cite{Aad:2015wra} and 13\TeV~\cite{Aaboud:2017cxo}, and by the CMS Collaboration at 8\TeV~\cite{Khachatryan:2015lba} and 13\TeV~\cite{Sirunyan:2019xls}.

\section{The CMS detector}

The central feature of the CMS apparatus is a superconducting solenoid
of 6\unit{m} internal diameter, providing a magnetic field of 3.8\unit{T}.
Within the solenoid volume are a silicon pixel and strip tracker,
a lead tungstate crystal electromagnetic calorimeter (ECAL),
and a brass and scintillator hadron calorimeter,
each composed of a barrel and two endcap sections.
Forward calorimeters extend the pseudorapidity ($\eta$) coverage
provided by the barrel and endcap detectors.
Muons are detected in gas-ionisation chambers embedded in
the steel flux-return yoke outside the solenoid.
A two-level trigger system~\cite{Khachatryan:2016bia} is used to reduce the rate of recorded events to a level suitable for data acquisition and storage.
A more detailed description of the CMS detector,
together with a definition of the coordinate system used
and the relevant kinematic variables, can be found in Ref.~\cite{cms}.

\section{Event simulation and background predictions}\label{sec:simulation}

Background samples for this search are produced for $\PZ$ boson production through the Drell--Yan (DY) process, top quark pair production (\ttbar{}), single top quark, diboson, triboson, $\ttbar\PV$ (${\PV = \PW,\ \PZ}$), $\PW$+jets, and SM Higgs boson production. They are generated at next-to-leading order (NLO) precision in perturbative quantum chromodynamics (QCD). In particular, the DY, $\ttbar\PV$, $\PW$+jets, triboson, and part of the diboson background samples are produced with \MGvATNLO versions~2.2.2~\cite{Alwall:2014hca} with the FxFx~\cite{Frederix:2012ps} procedure for NLO jet merging and \textsc{MadSpin}~\cite{Artoisenet:2012st} to properly propagate spin information in the matrix element of the process. The \ttbar, single top quark, SM Higgs boson production, and the remaining diboson background samples are produced with \POWHEG version~2~\cite{Nason:2004rx,Frixione:2007vw,Alioli:2010xd,Powheg_tt,Powheg_st}. The event generators are interfaced with \PYTHIA~8.212~\cite{Sjostrand:2014zea} for parton shower and hadronisation. The underlying event tune is CUEPT8M1~\cite{Khachatryan:2110213}, derived from the MONASH tune~\cite{Skands:2014pea}. The NNPDF 3.0~\cite{Ball:2014uwa} LO and NLO parton distribution functions (PDF) are used.

Signal samples of 207 different mass hypotheses are produced for the process $\PH \to \PZ\PSA \to \ell\ell\bbbar$,
with \mH and \mA ranging from 120 to 1000\GeV and from 30 to 1000\GeV, respectively. The choice of the mass hypotheses is strongly motivated by the need of achieving a complete coverage of the parameter space.
The spacing between two adjacent mass hypotheses is chosen so as to take into account the worsening of the signal resolution as the mass increases, such that the signal shape can be interpolated with good accuracy over the whole search region. These samples
are produced using \MGvATNLO version 2.3.2~\cite{Alwall:2014hca}
interfaced with \PYTHIA~8.212~\cite{Sjostrand:2014zea} for simulation of the parton shower, hadronisation, and underlying event using the CUEPT8M1 tune~\cite{Khachatryan:2110213}. The PDF set used is NNPDF~3.0~\cite{Ball:2014uwa} at leading order (LO) in the four-flavour scheme, and the
factorisation and renormalisation scales are estimated
dynamically. The total widths assumed for the \PH and \PSA resonances are computed with 2HDMC~\cite{Eriksson:2009ws}. Finally, the signal simulation does not account for production of the scalar \PH in association with {\cPqb} jets.

For all processes, the detector response is simulated using a detailed
description of the CMS apparatus, based on the \GEANTfour
package~\cite{Agostinelli2003250}. Additional $\Pp\Pp$ interactions in
the same and or neighbouring bunch crossings (pileup) are
generated with \PYTHIA~8.212~\cite{Sjostrand:2014zea}, and overlapped with the simulated events of
interest in order to reproduce the pileup measured in data.

All background processes are normalised to their most accurate
theoretical cross sections. The \ttbar, DY, single top quark,
$\PW^+ \PW^-$, and $\PW$+jets samples are normalised to next-to-next-to-leading order (NNLO) precision in
QCD~\cite{Czakon:2011xx,Li:2012wna,Butterworth:2015oua,Gehrmann:2014fva},
while the remaining diboson, triboson and $\ttbar\PV$ processes are
normalised to NLO precision in
QCD~\cite{Campbell:2011bn,Alwall:2014hca}. The SM Higgs boson production cross section is computed at NNLO QCD precision
and NLO electroweak precision~\cite{deFlorian:2016spz}. We indicate the SM Higgs boson, the $\ttbar\PV$, and the $\PW$+jets backgrounds with \textit{Other} in the figures.

\section{Event reconstruction and selection}\label{sec:event_reco}

Events considered for this search are selected by a trigger based on the dilepton signature. The
leading and subleading transverse momentum ($\pt$) thresholds applied by the triggers are
channel dependent, and vary from 17 to 23\GeV (8 to 12\GeV) for the leading (subleading) lepton. Trigger efficiencies
are measured with a ``tag-and-probe'' method~\cite{Chatrchyan2011}
as a function of lepton $\pt$ and $\eta$ in a data control region
consisting of $\PZ \to \ell\ell$ events. Events with two oppositely
charged leptons ($\Pe^{\pm}{}\Pe^{\mp}$, $\PGm^{\pm}{}\PGm^{\mp}$) are selected using asymmetric
$\pt$ requirements, chosen to be above the corresponding trigger
thresholds, for the leading and subleading leptons. These requirements are 25 and 15\GeV, respectively,
for $\Pe^{\pm}{}\Pe^{\mp}$ events; and 20 and 10\GeV, respectively, for $\PGm^{\pm}{}\PGm^{\mp}$ events. Electrons in the range $\abs{\eta} < 2.5$ and muons in the range $\abs{\eta} < 2.4$ are considered. Events with
different-flavour leptons ($\Pe^{\pm}{}\PGm^{\mp}$) are also selected. The $\pt$ requirement for the leading lepton is 25 and 15 (10)\GeV for the subleading electron (muon). These events mostly arise from \ttbar production, and this region is used in the final template fit described in Section~\ref{sec:results} to obtain an estimate of the normalisation of the non-resonant background processes (\ttbar{}, single top quark, diboson, and triboson) and of the shape of the \ttbar process only.
For simplicity, we will refer to events with two electrons, muons, and mixed-flavour leptons as $\Pe{}\Pe$, $\PGm{}\PGm$, and $\Pe{}\PGm$, respectively, throughout this paper.

A particle-flow (PF) algorithm~\cite{Sirunyan:2017ulk} aims at reconstructing all particles (PF candidates) in an event by combining information from all subdetectors. The PF candidates include photons, electrons, muons, neutral hadrons, and charged hadrons. The candidate vertex with the largest value of summed physics-object $\pt^2$ is taken to be the primary $\Pp\Pp$ interaction vertex. The physics objects used in this determination are the jets, clustered using the jet finding algorithm~\cite{Cacciari:2008gp,Cacciari:2011ma} with the tracks assigned to candidate vertices as inputs, and the associated missing transverse momentum, taken as the negative vector sum of the $\pt$ of those jets.
Electrons, reconstructed by associating tracks with ECAL clusters, are
identified by a sequential selection using information on the cluster
shape in the ECAL, track quality, and the matching between the track
and the ECAL cluster. Additionally, electrons from photon conversions
are rejected~\cite{Khachatryan:2015hwa}. Muons are reconstructed from
tracks found in the muon system, associated with tracks in the silicon
tracking detectors. They are identified based on the quality of the
track fit and the number of associated hits in the various tracking
detectors~\cite{Sirunyan:2018fpa}. The Rochester correction~\cite{Bodek:2012id} is applied to the muon momenta to correct for misalignments of the detector or uncertainties in the magnetic field.
The lepton isolation, defined as
the scalar \pt sum of all PF candidates in a cone of radius $\Delta R = 0.4$ around the
lepton, excluding the lepton itself and corrected for contributions from particles not coming from the primary vertex, divided by the lepton \pt, is
required to be $<$0.06 for electrons and
$<$0.15 for muons. Here, $\Delta R$ is defined in terms of the track separation
in $\eta$ and azimuthal angle ($\phi$, in radians) as $\Delta R = \sqrt{\smash[b]{(\Delta \phi)^2 + (\Delta \eta)^2}}$.
Moreover, the lepton tracks are required to be connected
to the primary vertex. Lepton
identification and isolation efficiencies in the simulation are
corrected for residual differences with respect to data. These
corrections are measured in a data sample enriched in $\PZ \to
\ell\ell$ events, using the ``tag-and-probe'' method, and are
parameterised as a function of lepton \pt and $\eta$.

Jet reconstruction is performed by clustering the PF candidates
to form jets using the anti-\kt clustering
algorithm~\cite{Cacciari:2008gp} with a distance parameter of 0.4,
implemented in the \FASTJET package~\cite{Cacciari:2011ma}. Jet
energies are corrected for residual nonuniformity and nonlinearity of
the detector response~\cite{Khachatryan:2016kdb}. Jets are required to
have $\pt > 20$\GeV, $\abs{\eta} < 2.4$, and be separated from
identified leptons by a distance $\Delta R > 0.3$.  The missing
transverse momentum vector, defined as the projection onto the
transverse plane relative to the beam axis, of the negative vector sum
of the momenta of all PF candidates, is referred to as
\ptvecmiss~\cite{metPF,Sirunyan:2019kia}. Its magnitude is denoted
by \ptmiss. Corrections to the jet energies are propagated to
\ptvecmiss.

The DeepCSV algorithm~\cite{Sirunyan:2017ezt} is used to
identify jets originating from {\cPqb} quarks. Jets are considered as {\cPqb} tagged if they have $\pt > 20$\GeV and they pass the medium working point of the algorithm, which
provides around 70\% efficiency with a mistag rate of less than
1\%, while the mistag rate for {\cPqc} jets is around 10\%. Correction factors are applied in the simulation to the selected
jets to account for the different response of the DeepCSV algorithm
between data and simulation~\cite{Sirunyan:2017ezt}. Among all
possible dijet combinations fulfilling the previous criteria, we
select the two jets with the highest DeepCSV algorithm outputs.

The final object selection consists of two opposite-sign leptons and two {\cPqb}-tagged jets, after which a requirement of $70 < \mll
< 110$\GeV is applied to enhance the presence of $\PZ \to \ell\ell$
events. In addition, the events are required to have a $\ptmiss <
80$\GeV in order to reduce the background contributions from processes
with large $\ptmiss$, such as \ttbar production. Both requirements have negligible impacts on the signal efficiency.

The main background processes, in decreasing order of importance, are DY in association with {\cPqb} quarks and \ttbar production where both top quarks decay leptonically (fully leptonic \ttbar{}). The contribution from QCD multijet events with jets misidentified as leptons constitutes a negligible background after requiring a pair of well-identified leptons, as described in Section~\ref{sec:event_reco}.

\section{Signal extraction}\label{sec:strategy}

\begin{figure*}[!htb]
  \centering
  \includegraphics[width=0.48\textwidth]{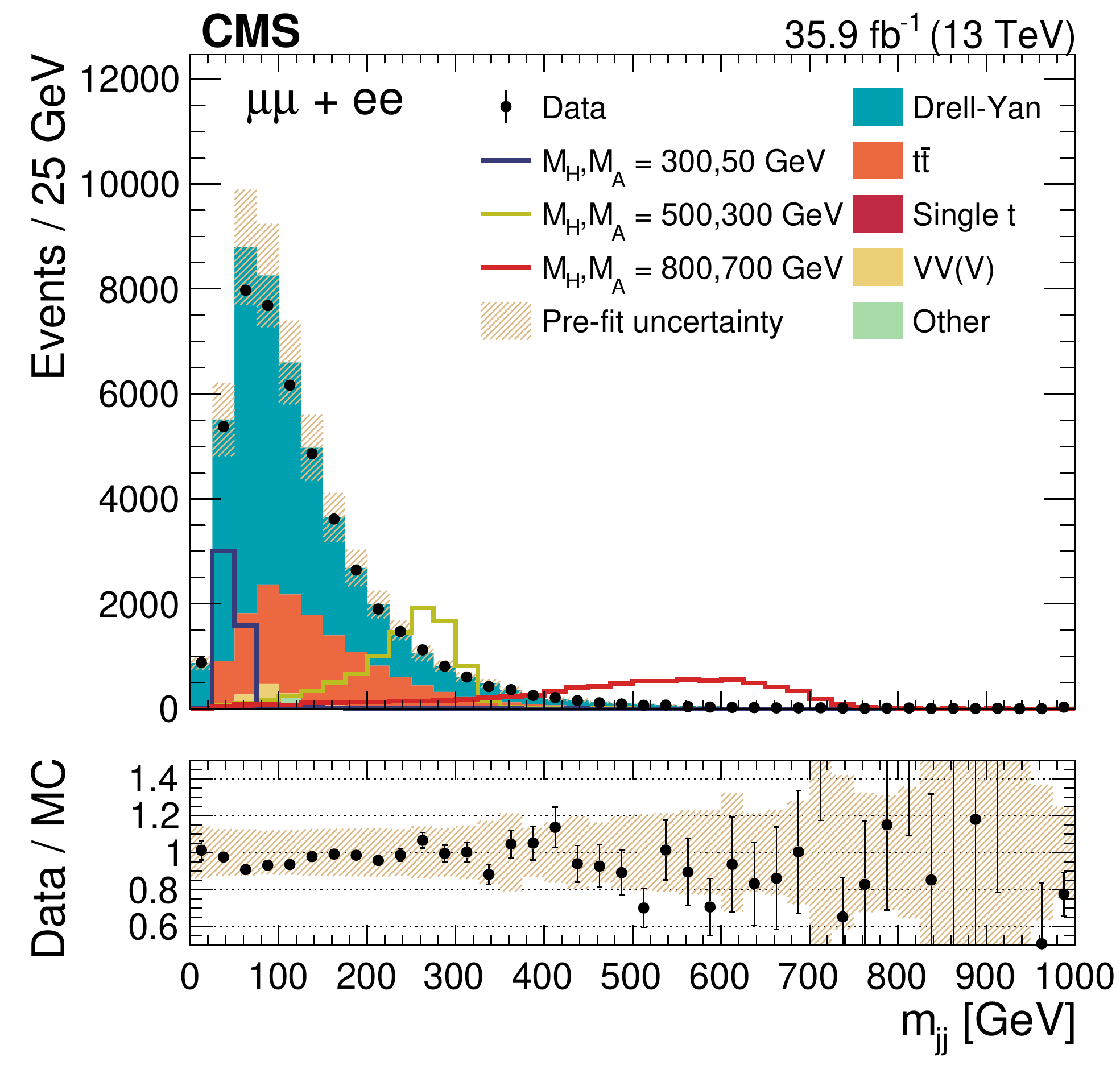}
  \includegraphics[width=0.48\textwidth]{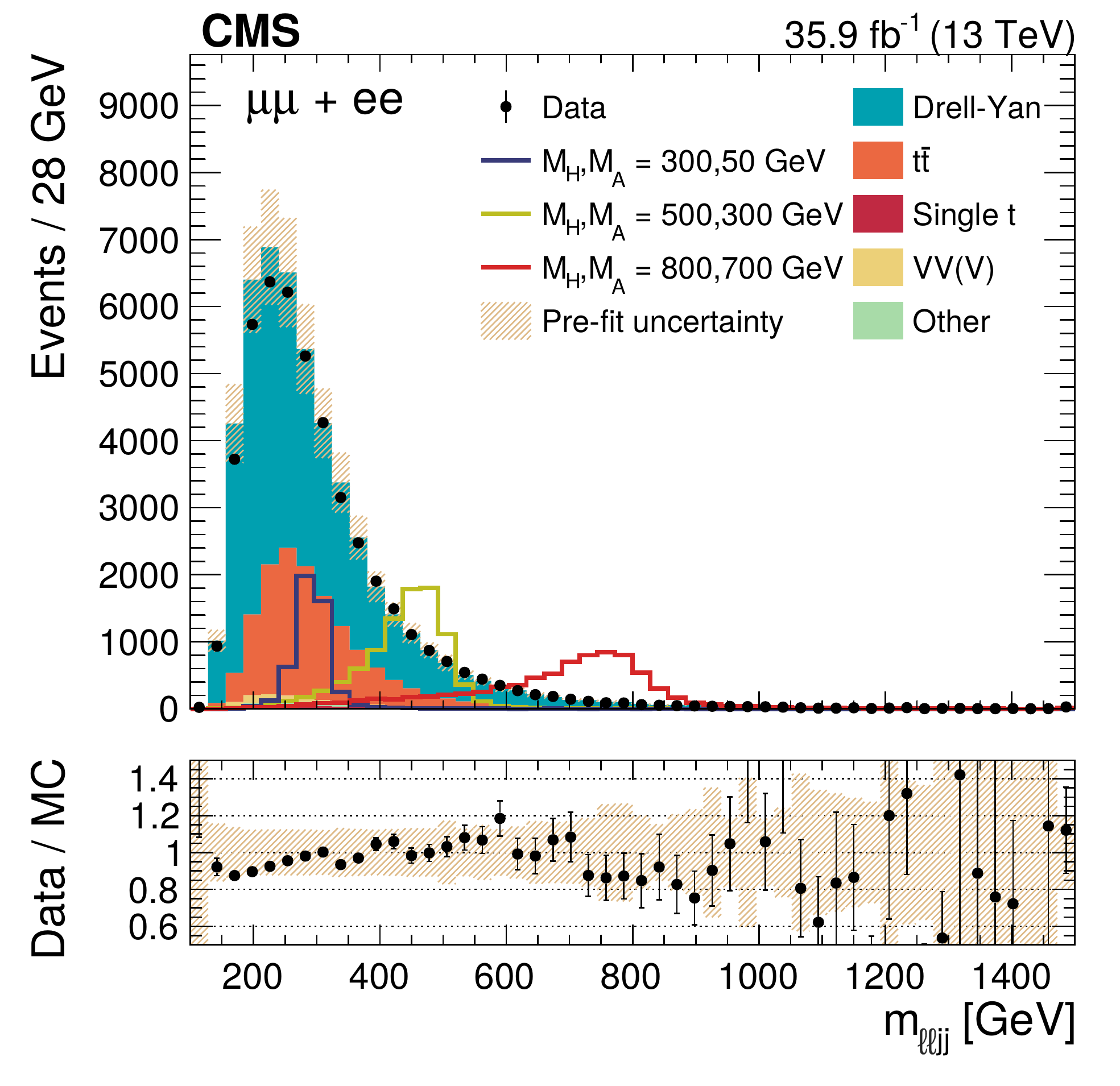}
    \caption{
    The $\mjj$ (left) and $\mlljj$ (right) distributions in data and background events after requiring all the analysis selections, for $\PGm{}\PGm + \Pe{}\Pe$ events. The background shapes and normalisations are obtained from simulation. The various signal hypotheses displayed have been scaled to a cross section of 2\unit{pb} for illustrative purposes. Error bars indicate statistical uncertainties, while shaded bands show systematic uncertainties prior to the fit (introduced in Section~\ref{sec:systematics}).
  }
    \label{fig:masses}
\end{figure*}

We search for the process $\PH \to \PZ\PSA \to \ell\ell\bbbar$ by
fully reconstructing its final-state objects and applying selection requirements in order to remove
as many background events as possible, as explained in Section~\ref{sec:event_reco}.
From the reconstructed objects, we search for resonances in the invariant masses. Specifically, the invariant mass of the \PSA can be reconstructed from the {\cPqb} jet pair; and that of the \PH from the {\cPqb} jet pair and the lepton pair.
Two categories are defined based on the lepton flavours considered: $\Pe{}\Pe$ and $\PGm{}\PGm$. The \PZ
mass, reconstructed from two opposite-sign leptons, is used in the selection
criteria described in Section~\ref{sec:event_reco} since it is common to
all signals studied in this paper. The masses of the other two particles, \PH and \PSA, vary according
to the signal scenarios considered. Therefore, a simple and
effective model independent approach to isolate the signal is to search for an excess of events in the reconstructed \mjj and \mlljj
distributions centered around the \PH and \PSA candidate mass for each signal hypothesis. These distributions for $\PGm{}\PGm + \Pe{}\Pe$ events are shown in Fig.~\ref{fig:masses}, where the background shapes and normalisations are obtained from simulation.

\begin{figure*}[!htb]
  \centering
    \includegraphics[height=0.4\textwidth]{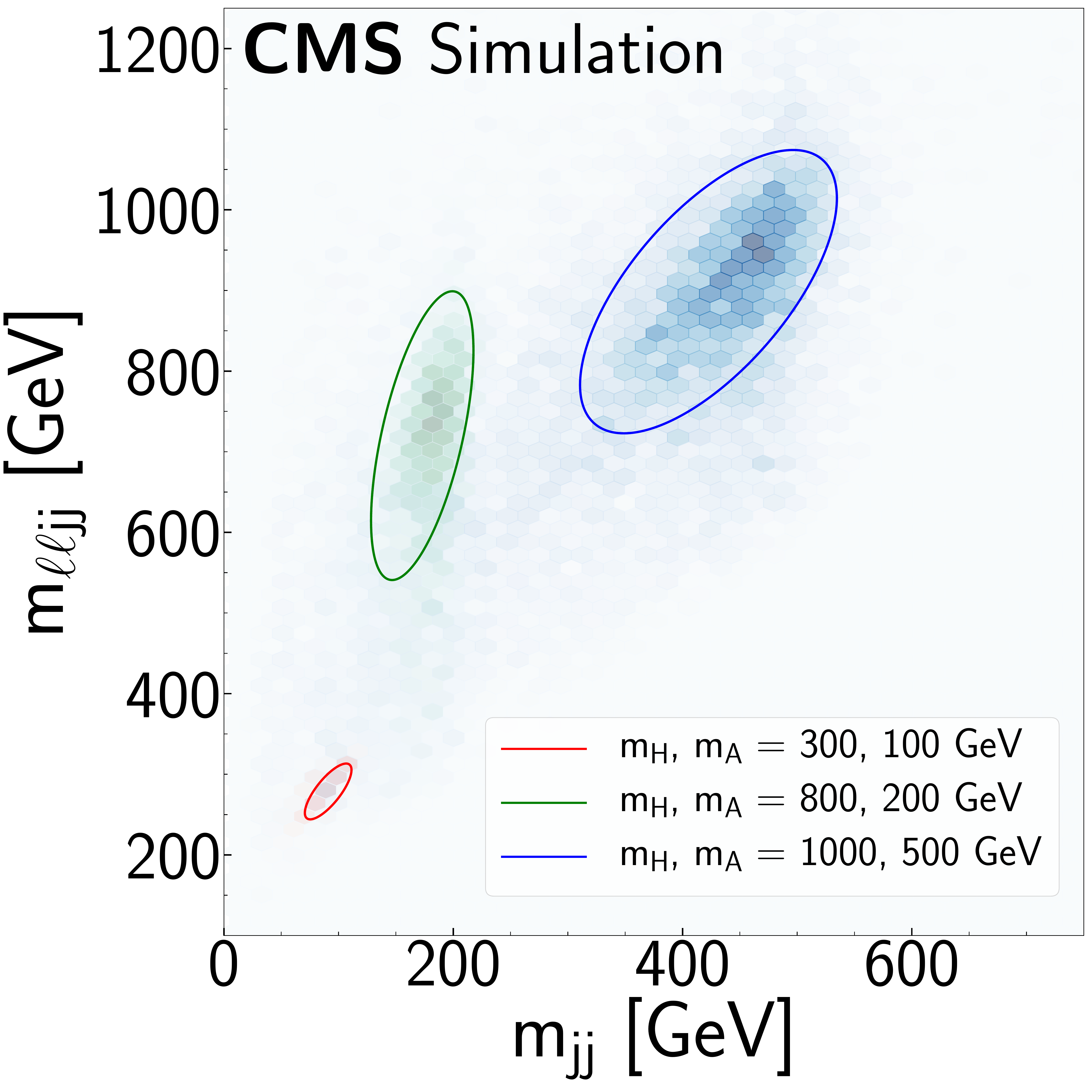}
    \includegraphics[height=0.47\textwidth]{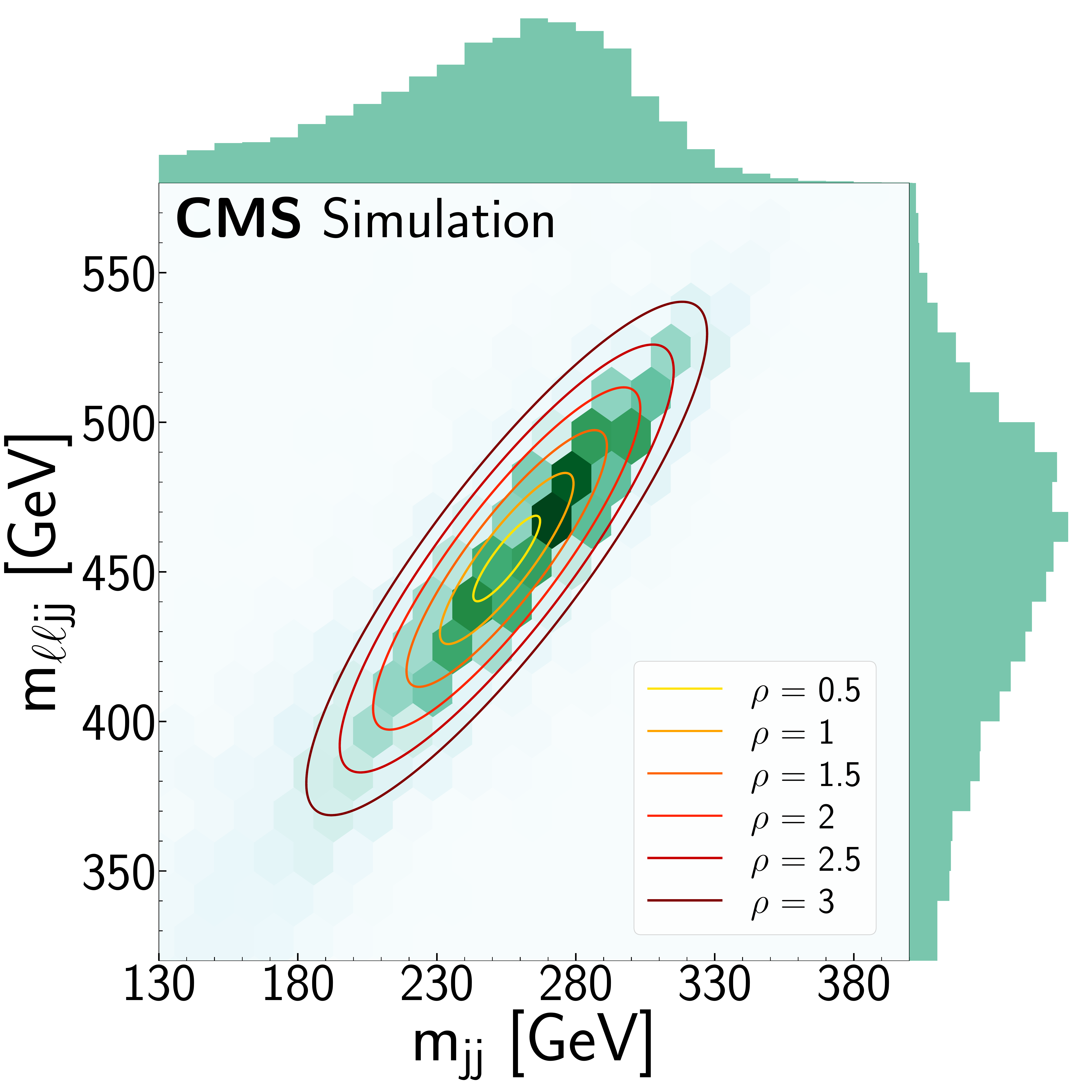}
    \caption{
     The $\mlljj$~vs.~$\mjj$ plane for signal samples under three different mass hypotheses, on which the parametrised ellipses are shown (left). A signal hypothesis with ${\mH=500}$\GeV and ${\mA=300}$\GeV is shown in the $\mlljj$~vs.~$\mjj$ plane (right). The different ellipses show the variation of the $\rho$ parameter in steps of 0.5, from 0 to 3. The intensity of the colour in each hexagonal bin is proportional to the number of events in it.
    }
   \label{fig:ellipses}

\end{figure*}

Since the \mjj and \mlljj distributions are inherently positively correlated under a particular signal hypothesis, an elliptical signal region is chosen in order to optimize the sensitivity of the search.
Figure~\ref{fig:ellipses} (left) shows the reconstructed mass distributions for three different signals in the $\mlljj$~vs.~$\mjj$ plane along with their defined elliptical signal regions. Because the shape of the signal is driven by the energy resolution of the final-state objects, ellipses take different sizes and tilt angles, depending on the masses being considered.
  A parametrisation is therefore performed in order to guarantee a good description of the signal shape for each signal hypothesis. For each ellipse, it provides the center, the major and minor semi-axes, and the tilt angle.
Since each ellipse must be well-centered around the maximum of the two-dimensional (2D) mass distribution, the reconstructed center is extracted from a one-dimensional Gaussian fit in both \mjj and \mlljj. The diagonalisation of the covariance matrix of the 2D distribution provides the axes of the ellipse and its tilt angle.

Since the shape of the signal is not exactly Gaussian, concentric elliptically shaped regions are defined in the parameter space using a parameter called $\rho$. Specifically, an ellipse with $\rho=i$ contains roughly the fraction of signal events expected within $i$ standard deviations in a 2D distribution. Selected events in the $\mlljj$~vs.~$\mjj$ plane are classified in six regions around the center of the ellipse defined for each signal point. The regions are built in $\rho$ steps of 0.5, from 0 to 3, as illustrated in Fig.~\ref{fig:ellipses} (right), and lead to a template containing six bins used to perform the statistical analysis. By construction, the bulk of the signal is located at small values of $\rho$.
The yield in data and the expected yields in simulation are reported in Table~\ref{table:yields} for each elliptical bin under the mass hypothesis ${\mH=500}$\GeV and ${\mA=200}$\GeV. The \Pe{}\Pe and \PGm{}\PGm categories are summed.

\begin{table*}[htpb]
\topcaption{
  Expected and observed event yields prior to the fit in the signal region with ${\mH=500}$\GeV and ${\mA=200}$\GeV for each elliptical bin. The signal is normalised to its theoretical cross section for the Type-II 2HDM benchmark ${\tan\beta=1.5}$ and ${\cos (\beta - \alpha)=0.01}$. The \Pe{}\Pe and \PGm{}\PGm categories are summed. The reported uncertainties are statistical only.
}
\label{table:yields}
\centering
\cmsTable{
\begin{tabular}{@{}lcccccc@{}} \hline
Process & \multicolumn{6}{c}{Yield} \\
  & $0 \leq \rho < 0.5$ & $0.5 \leq \rho < 1$  & $1 \leq \rho < 1.5$  & $1.5 \leq \rho < 2$  & $2 \leq \rho < 2.5$  & $2.5 \leq \rho < 3$  \\
\hline
DY  &  181 $\pm$ 14 & 438 $\pm$ 22  & 607 $\pm$ 27 & 987 $\pm$ 34 & 1440 $\pm$ 42 & 2273 $\pm$ 53 \\
\ttbar & 166 $\pm$ 2 & 420 $\pm$ 4 & 603 $\pm$ 5 & 826 $\pm$ 5 & 1165 $\pm$ 6 & 1597 $\pm$ 8 \\
Single top quark & 2.2 $\pm$ 0.5 & 6.2 $\pm$ 0.8 & 9 $\pm$ 1 & 17 $\pm$ 1 & 25.5 $\pm$ 1.7 & 38 $\pm$ 2 \\
\PV\PV{}(\PV) & 0.6 $\pm$ 0.1 & 1.9 $\pm$ 0.2 & 2.5 $\pm$ 0.2 & 3.9 $\pm$ 0.5 & 5.2 $\pm$ 0.4 & 9.1 $\pm$ 0.4 \\
Other &  0.9 $\pm$ 0.2 & 3.7 $\pm$ 0.3 & 5.1 $\pm$ 0.3 & 8.4 $\pm$ 0.4 & 11.7 $\pm$ 0.5 & 18.1 $\pm$ 0.6 \\
Total bkg. &  351 $\pm$ 14 & 870 $\pm$ 22 & 1227 $\pm$ 27 & 1842 $\pm$ 34 & 2647 $\pm$ 42 & 3935 $\pm$ 54 \\
Data  & 365 & 854 & 1231 & 1834 & 2608 & 3906 \\
Signal & 71.5 $\pm$ 1.3 & 122.7 $\pm$ 1.7 & 86.1 $\pm$ 1.4 & 48.0 $\pm$ 1.0 & 26.6 $\pm$ 0.8 & 17.5 $\pm$ 0.6 \\
\hline
\end{tabular}
}
\end{table*}

\section{Systematic uncertainties} \label{sec:systematics}
We consider different sources of systematic uncertainties that may affect the statistical interpretation of the results, through their modification of both the normalisation and the shape of the distributions for the signal and background processes.

Theoretical uncertainties in the cross sections of the background processes
estimated using simulation are considered as systematic uncertainties
in the yield predictions. The uncertainty in the total integrated
luminosity is determined to be 2.5\%~\cite{CMS-PAS-LUM-17-001}.

The signal region contains events that have at least two {\cPqb}-tagged jets. A control region is built by requiring events to pass the selection, as described in Section~\ref{sec:event_reco}, but with no \Pb tag requirement for the jets. In that region, a discrepancy between data and simulation of up to 10\% is observed in the shape of the mass distributions,
which hints at a mismodeling of the DY~+~heavy-flavour jets background in some specific regions of the reconstructed mass plane.
To account for this mismodeling, given the similar precision of the DY~+~light-flavour and DY~+~heavy-flavour jets background samples at simulation level, the observed data-MC discrepancy is fitted with a polynomial function, which is used to reweight each DY~+~heavy-flavour jets simulated event in the signal region, and a shape uncertainty equal to 100\% of the correction is applied. In order to avoid assigning only one shape uncertainty to regions characterised by very different values of the above-mentioned correction, this uncertainty is considered independently in 42 regions of approximately $150\times150\GeV^2$ in the $\mlljj$~vs.~$\mjj$ plane. This procedure ensures enough degrees of freedom in the maximum likelihood fit (used to extract the best fit signal cross section, as explained in Section~\ref{sec:results}) to properly account for the mismodeling of the DY~+~heavy-flavour jets background shape.

The following sources of systematic uncertainties that affect the
normalisation and shape of the templates used in the statistical
evaluation are considered:
\begin{itemize}

\item \textit{Trigger efficiency, lepton identification and
isolation:} uncertainties in the measurement of trigger efficiencies, as well as
electron and muon isolation and identification efficiencies, are
considered. These are evaluated
as a function of lepton $\pt$ and $\eta$, and their effect on the
analysis is estimated by varying the corrections to the efficiencies by $\pm$1~standard deviation.

\item \textit{Jet energy scale and resolution:} uncertainties in the
jet energy scale are of the order of a few percent and are estimated as
a function of jet $\pt$ and $\eta$~\cite{Khachatryan:2016kdb}.  A
difference in the jet energy resolution of about 10\% between data and
simulation is accounted for by worsening the jet energy resolution in
simulation by $\eta$-dependent factors. The uncertainty due to these
corrections is estimated by a variation of the factors applied by
$\pm$1~standard deviation. Variations of jet energies are propagated
to $\ptvecmiss$.

\item \textit{{\cPqb} tagging:} {\cPqb} tagging efficiency and light-flavour
mistag rate corrections and associated uncertainties are determined as
a function of the jet $\pt$~\cite{Sirunyan:2017ezt}. Their effect on the
analysis is estimated by varying these corrections by $\pm$1~standard
deviation.

\item \textit{Pileup:} the measured total inelastic cross section is
varied by $\pm$4.6\%~\cite{Sirunyan:2018nqx} to produce different expected
pileup distributions.

\item \textit{Renormalisation and factorisation scale uncertainty:}
this uncertainty is estimated by varying the renormalisation
($\mu_{\mathrm{R}}$) and the factorisation ($\mu_{\mathrm{F}}$) scales
used during the generation of the simulated samples independently by
factors of 0.5, 1, or 2. Cases where the two scales are at
opposite extremes, are not considered.  An envelope is built from the
6 possible combinations by keeping maximum and minimum variations for
each bin of the distributions, and is used as an estimate of the scale
uncertainties for all the background and signal samples.

\item \textit{PDF uncertainty:} the magnitudes of the uncertainties
related to the PDFs and the variation of the strong coupling constant
for each simulated background and signal process are obtained using
variations of the NNPDF~3.0 set~\cite{Ball:2014uwa}, following the
PDF4LHC prescriptions~\cite{Butterworth:2015oua}.

\item \textit{Drell--Yan additional uncertainty:} additional shape uncertainties are applied to DY events to correct for mismodeling of this background as explained above. Their values range up to 10\%, depending on the region of the reconstructed mass plane.

\item \textit{Simulated sample size:} the finite nature of simulated
samples is considered as an additional source of systematic
uncertainty. For each bin of the distributions, one additional
uncertainty is added, where only the considered bin is altered by
$\pm$1~standard deviation, keeping the others at their nominal value.

\end{itemize}

The variations that these uncertainties induce on the total event yields in the analysis
selection are summarised in Table~\ref{table:sys} for a specific signal hypothesis, where the $\Pe{}\Pe$ and $\PGm{}\PGm$ categories are combined together, yielding, for some uncertainties, a range of variations.

\begin{table*}[htb]
\topcaption{
  Summary of the systematic uncertainties prior to the fit and the variation, in percentages, that they induce on the total event yields for the dominant background and signal processes, under a particular signal hypothesis with ${\mH = 379}$\GeV and ${\mA = 172}$\GeV.
}
\label{table:sys}
\centering

\begin{tabular}{@{}lcc@{}} \hline
Source & Background yield variation & Signal yield variation \\
\hline
Electron identification and isolation   & 2.7\%  & 2.6\% \\
Integrated luminosity   & 2.5\%  & 2.5\% \\
Jet energy scale   &  2.1--2.4\%   &   0.1--0.3\% \\
\Pb tagging (heavy-flavour jets)   & 2.3\%  & 2.0\% \\
PDFs   &  1.0\%   &   0.5\% \\
Pileup   & 0.3--0.9\%  & 0.7--1.3\% \\
\Pb tagging (light-flavour jets)   & 0.7--0.8\%  & $<$0.1\% \\
Muon identification and isolation   & 0.5\%  & 0.4\% \\
Trigger efficiency   & 0.1--0.3\%  & 0.1--0.3\% \\
Jet energy resolution   & 0.2\%  & 0.2\% \\
\multicolumn{3}{c}{Affecting only \ttbar (31.8\% of the total bkg.)}\\
$\mu_\mathrm{R}$ and $\mu_\mathrm{F}$ scales   & 12.2--12.3\% &  \\
\ttbar cross section &  5.3\%  &  \\
\multicolumn{3}{c}{Affecting only Drell--Yan (64.5\% of the total bkg.)}\\
$\mu_\mathrm{R}$ and $\mu_\mathrm{F}$ scales   & 9.6\%   & \\
Drell--Yan cross section   & 4.9\%  & \\
Drell--Yan additional uncertainty   & 2.1--2.2\%  & \\
Simulated sample size   & 0.5--1.3\%  & \\
\multicolumn{3}{c}{Affecting only \PV\PV (1.1\% of the total bkg.)}\\
$\mu_\mathrm{R}$ and $\mu_\mathrm{F}$ scales   & 4.3--4.8\%  & \\
\multicolumn{3}{c}{Affecting only signal} \\
$\mu_\mathrm{R}$ and $\mu_\mathrm{F}$ scales   &   &  1.8\%\\
\hline
\end{tabular}

\end{table*}

\section{Methodology and results}
\label{sec:results}

A binned maximum likelihood fit is performed in order to extract best fit
signal cross sections. The fit is performed using the six binned
templates mentioned above in the \Pe{}\Pe and \PGm{}\PGm channels. An additional six-bin template is included in the fit containing the $\Pe\PGm$ selection to further constrain the \ttbar process, which is the major background component in this region, together with the minor non-resonant background processes. The systematic uncertainties are introduced as nuisance parameters in the fit. For each systematic uncertainty affecting the shape (normalisation) of the templates, a nuisance parameter is constrained with a Gaussian (log-normal) prior. The best fit values for all the nuisance parameters, as well as the
corresponding uncertainties, are extracted by performing a
binned maximum likelihood fit to
the data.

\begin{figure}[b]
  \centering
  \includegraphics[width=0.49\textwidth]{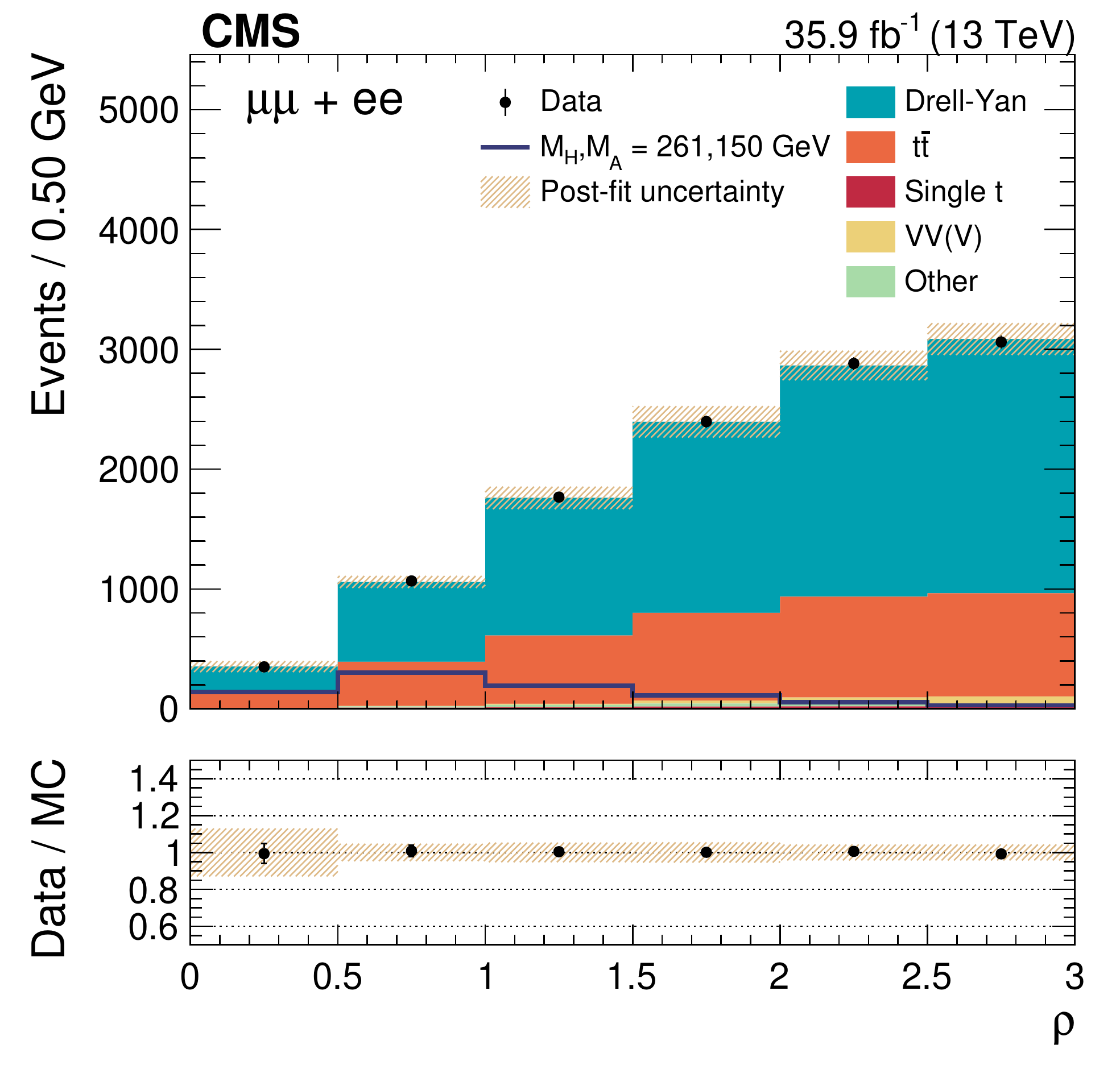}
  \includegraphics[width=0.49\textwidth]{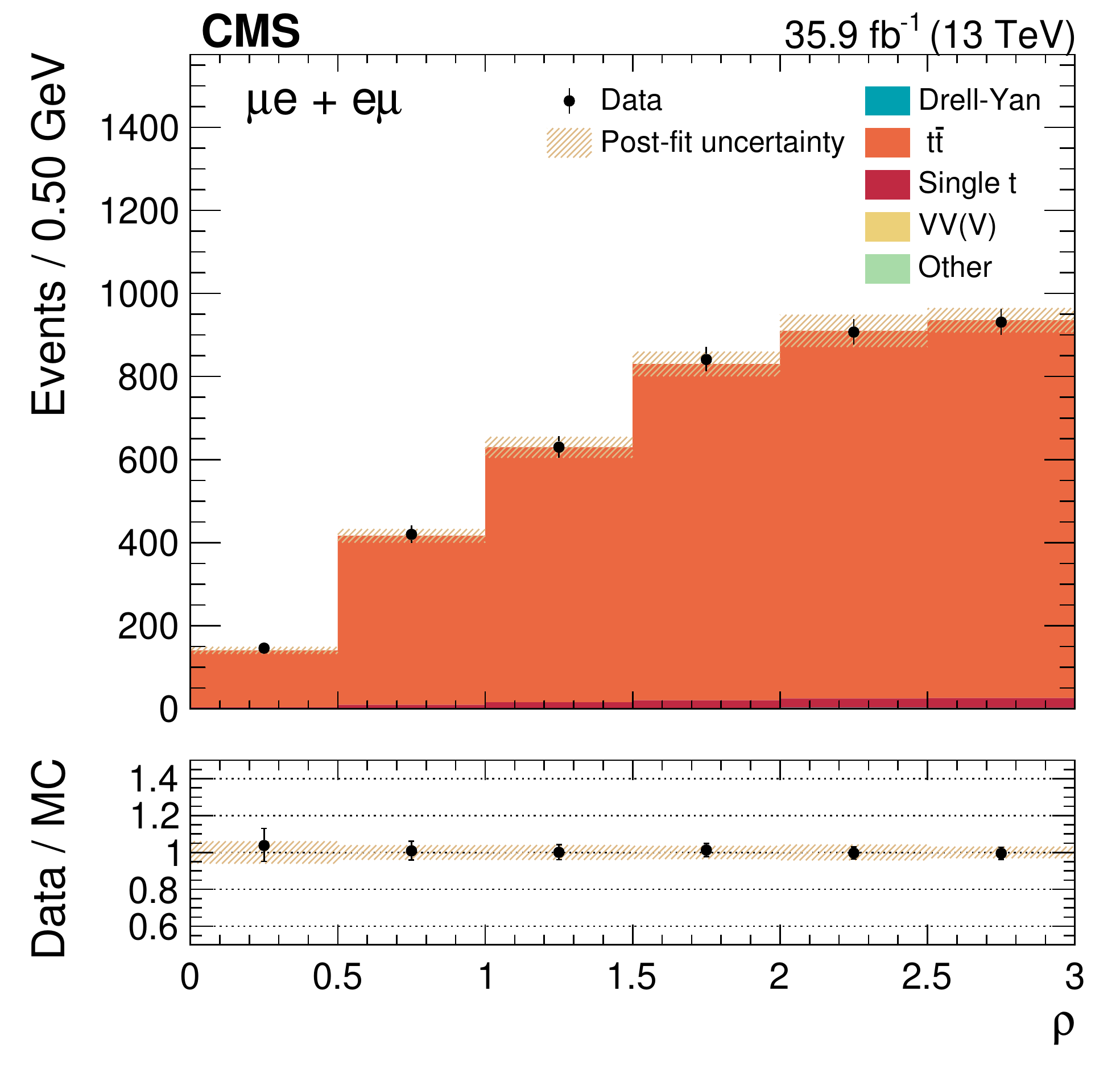}
  \includegraphics[width=0.49\textwidth]{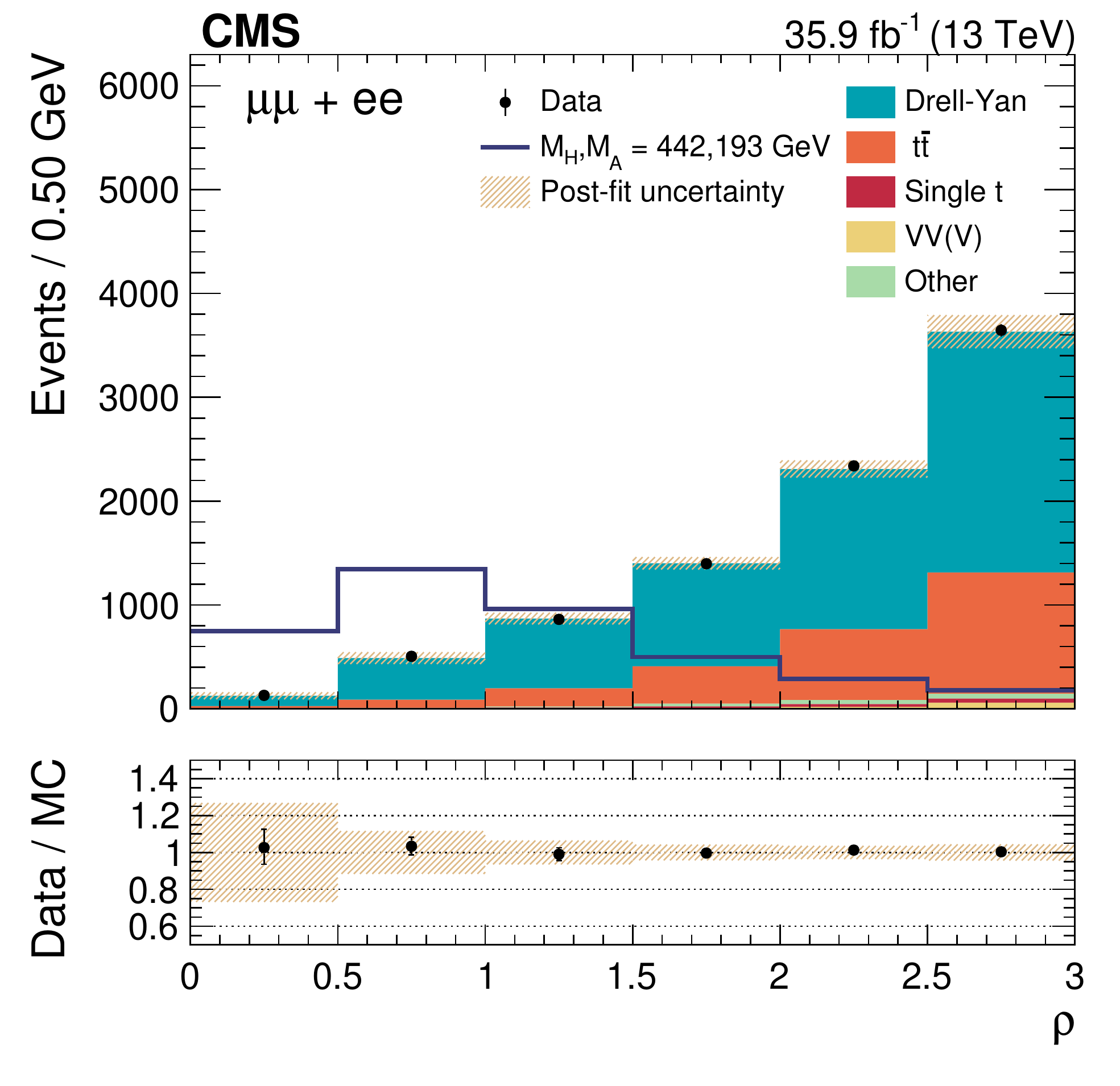}
  \includegraphics[width=0.49\textwidth]{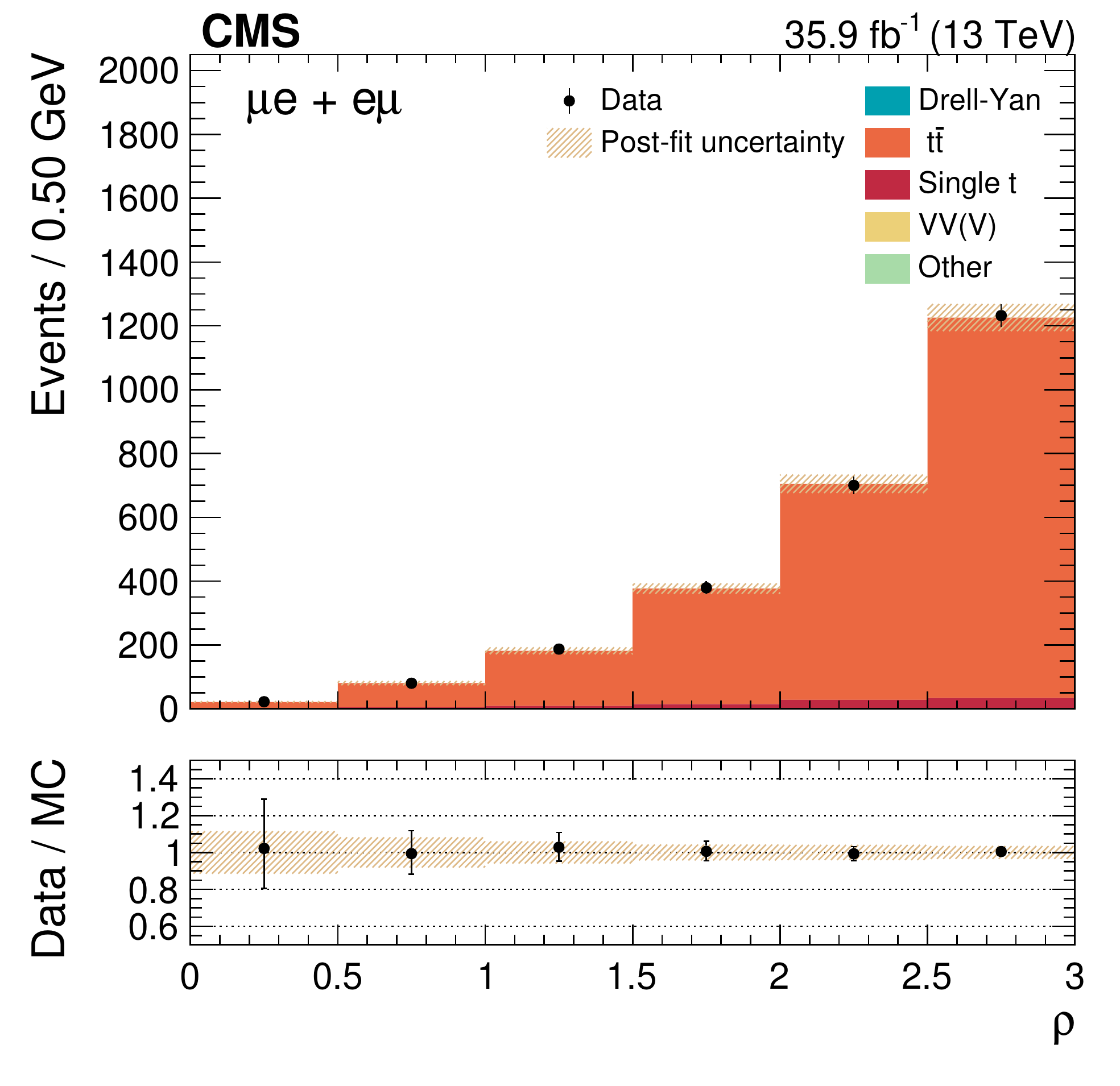}
  \caption{
       Post-fit $\rho$ distributions from a background-only fit for the same-flavour lepton (left) and mixed-flavour lepton (right) events corresponding to a signal hypothesis with ${\mH=261}$\GeV and ${\mA=150}$\GeV (upper) and ${\mH=442}$\GeV and ${\mA=193}$\GeV (lower). The signal is normalised to 1\unit{pb}. Error bars indicate statistical
      uncertainties, while shaded bands show systematic uncertainties after the fit.
  }
    \label{fig:final_plot_1}
\end{figure}

Figure~\ref{fig:final_plot_1} shows final distributions of events after a background-only fit in bins of $\rho$ under two different mass hypotheses for the $\PGm{}\PGm + \Pe{}\Pe$ and $\Pe\PGm$ categories with all the nuisance parameters set to their best fit values. The corresponding signals are also displayed and normalised to 1\unit{pb} for illustrative purposes.

\begin{figure}[tb]
  \centering
  \includegraphics[width=0.6\textwidth]{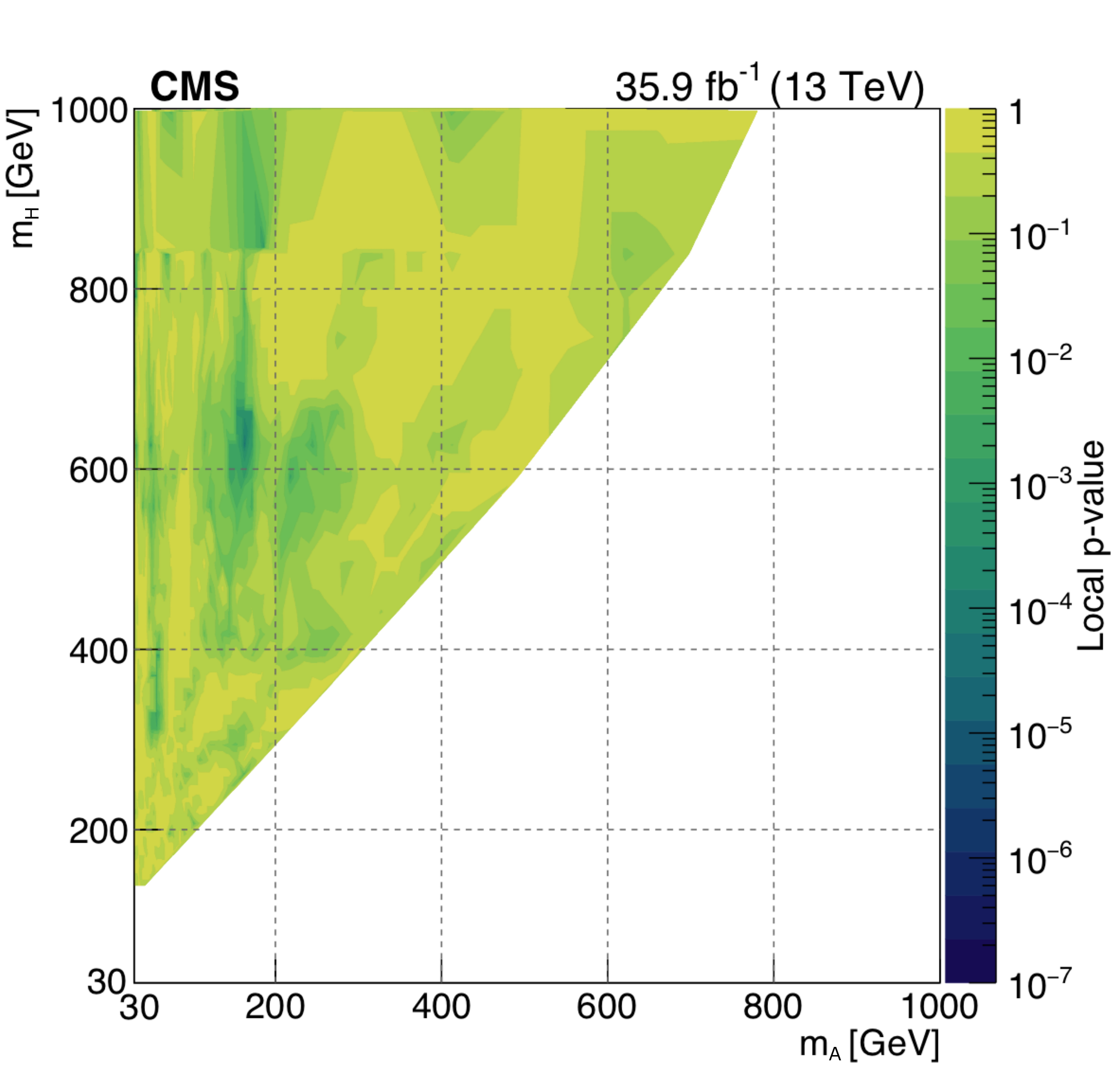}
  \caption{
    Distribution of the local p-value in the $\mH$~vs.~$\mA$ plane.
  }
    \label{fig:pval}
\end{figure}

No significant deviations from the SM expectations are observed. The highest asymptotic local significance observed corresponds to 3.9~standard deviations for the signal hypothesis with ${\mH=627}$\GeV and ${\mA=162}$\GeV, which globally becomes 1.3~standard deviations once accounting for the look elsewhere effect~\cite{Gross:2010qma}, evaluated with the method described in Ref.~\cite{Vitells:2011da}. The local p-value in the $\mH$~vs.~$\mA$ plane is displayed in Fig.~\ref{fig:pval}.

\begin{figure}[tb]
  \centering
  \includegraphics[width=0.49\textwidth]{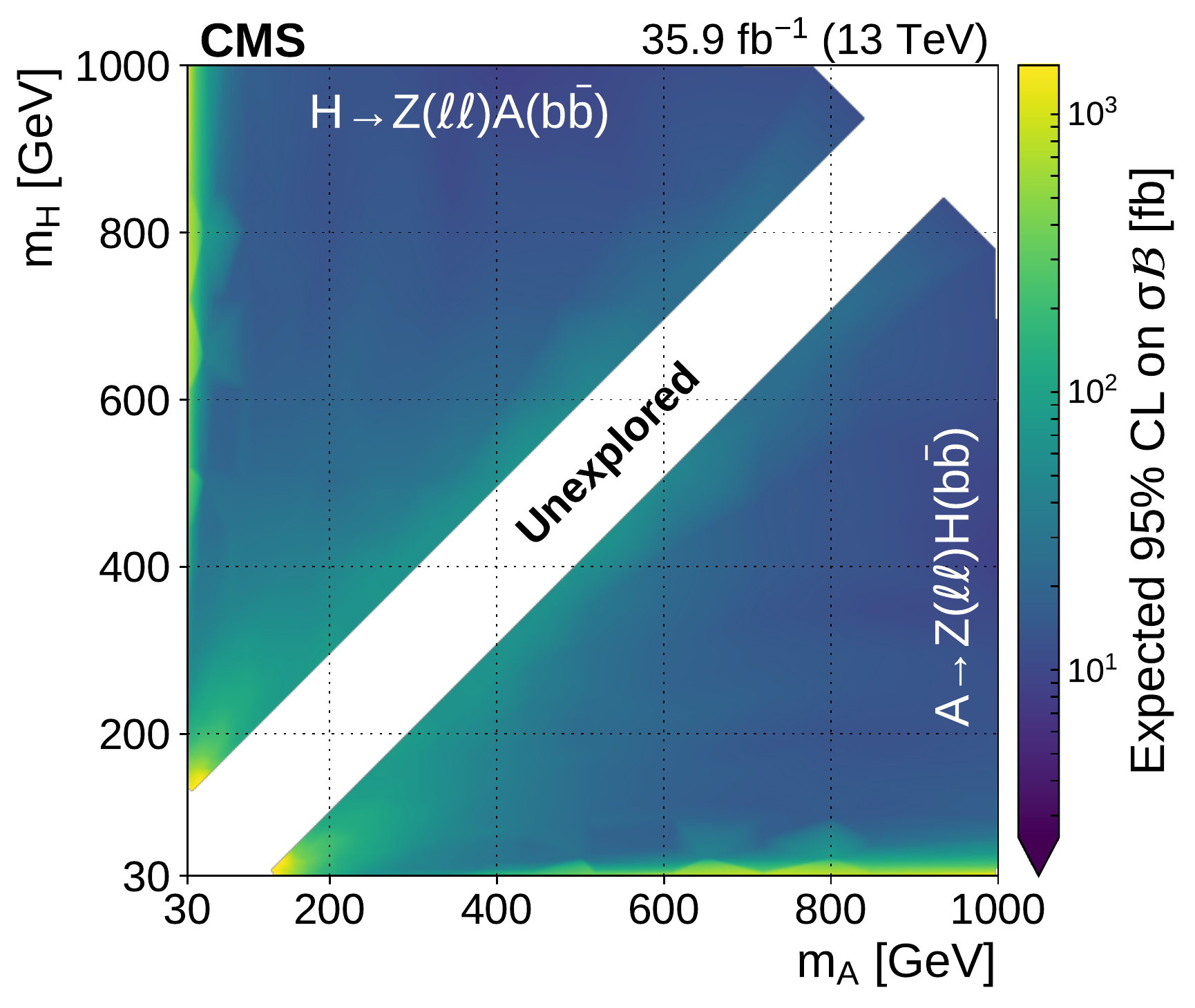}
  \includegraphics[width=0.49\textwidth]{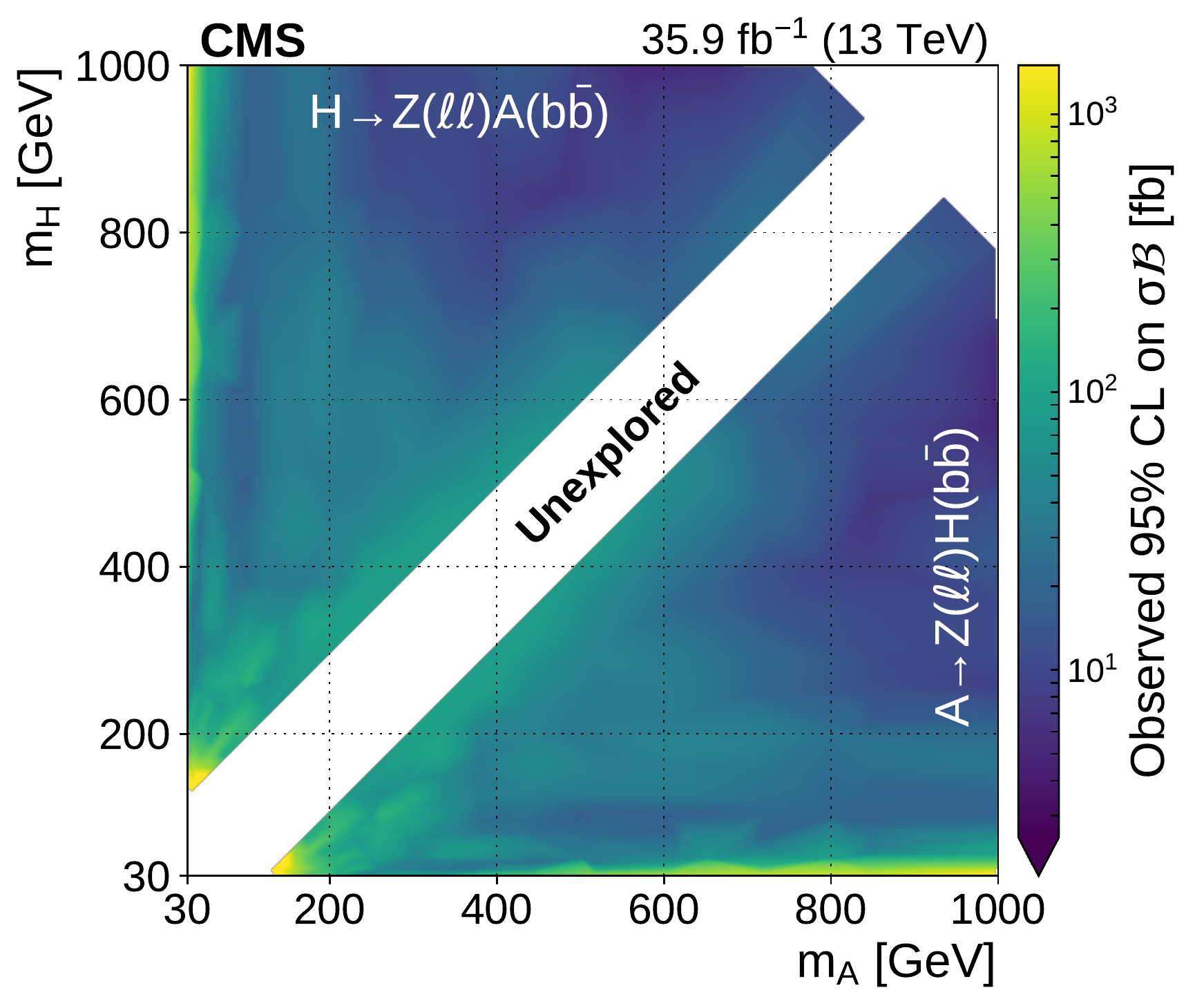}
  \caption{
    Expected (left) and observed (right) 95\% \CL upper limits on the product of the production cross section and branching fraction $\sigma \mathcal{B}$ for $\PH(\PSA) \to\PZ\PSA(\PH) \to \ell\ell\bbbar$ as a function of \mA and \mH. The limits are computed using the asymptotic \CLs method, combining the \Pe{}\Pe and \PGm{}\PGm channels.
  }
    \label{fig:observed}
\end{figure}

\begin{figure}[tb]
  \centering
  \includegraphics[width=0.58\textwidth]{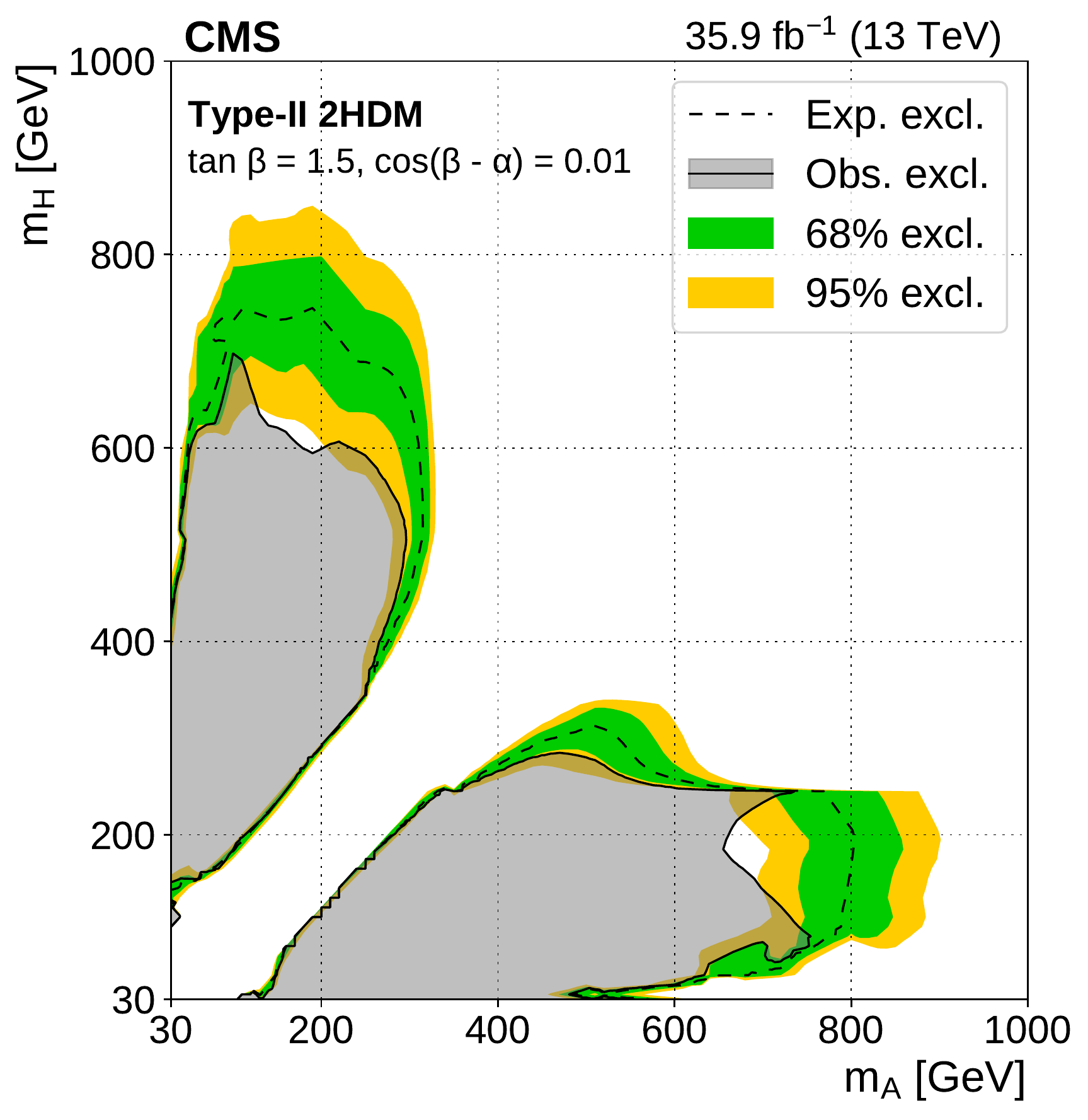}
  \caption{
    Expected and observed 95\% \CL exclusion contours for the Type-II 2HDM benchmark ${\tan\beta=1.5}$ and ${\cos (\beta-\alpha)=0.01}$ as a function of \mA and \mH. The inner (green) band and the outer (yellow) band indicate the regions containing 68 and 95\%, respectively, of the distribution of the exclusion contours expected under the background-only hypothesis. The limits are computed using the asymptotic \CLs method, combining the \Pe{}\Pe and \PGm{}\PGm channels.
  }
    \label{fig:limits_2D}
\end{figure}

\begin{figure}[tb]
  \centering
  \includegraphics[width=0.58\textwidth]{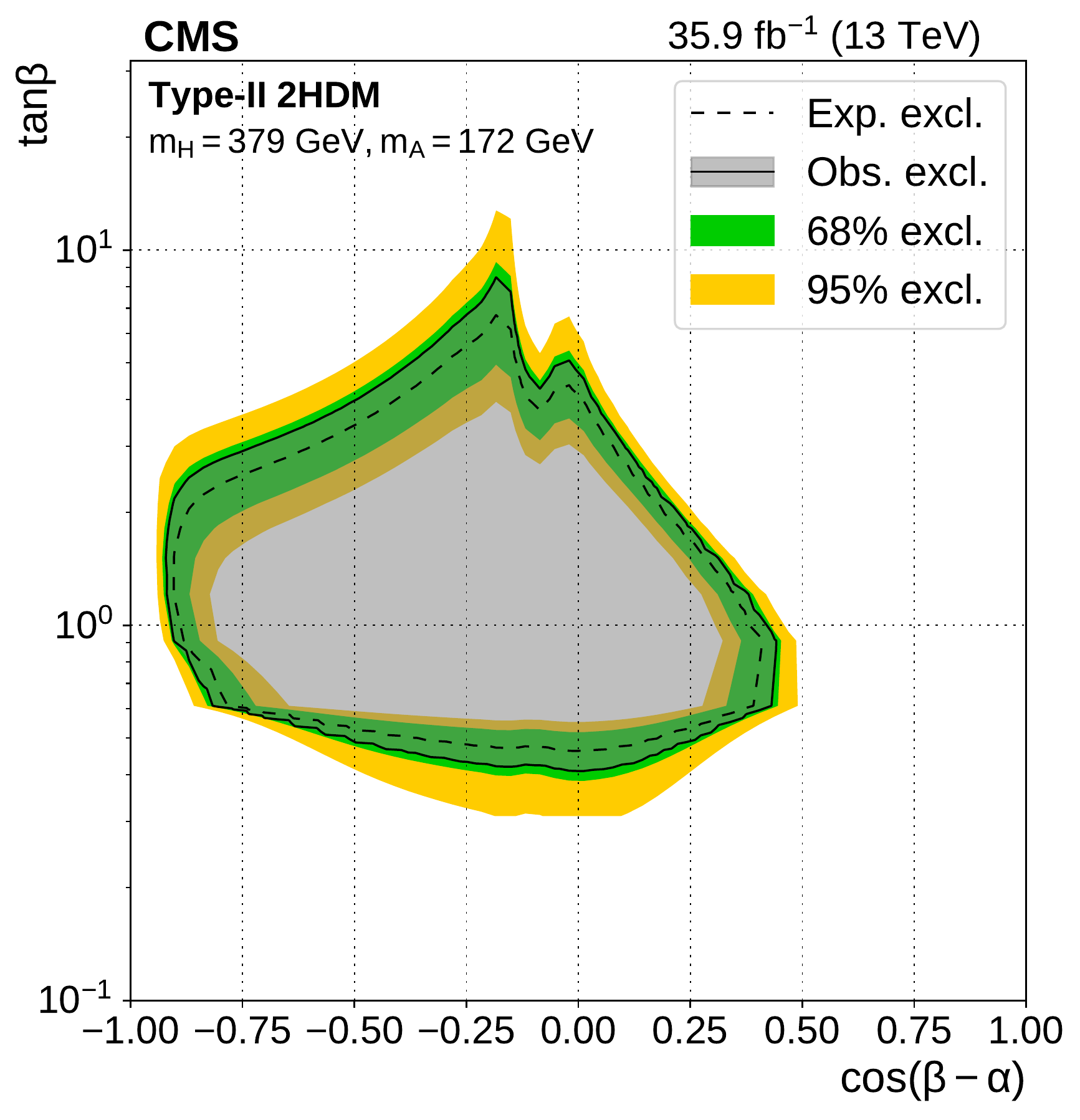}
  \caption{
    Expected and observed 95\% \CL exclusion contours for ${\mH=379}$\GeV and ${\mA=172}$\GeV as a function of $\tan \beta$ and ${\cos (\beta - \alpha)}$. The inner (green) band and the outer (yellow) band indicate the regions containing 68 and 95\%, respectively, of the distribution of the exclusion contours expected under the background-only hypothesis. The limits are computed using the asymptotic \CLs method, combining the \Pe{}\Pe and \PGm{}\PGm channels.
  }
    \label{fig:limits_2D_tb_cba}
\end{figure}

Figure~\ref{fig:observed} shows expected (left) and observed (right) model independent upper limits at 95\% confidence level (\CL), ${\sigma_{95\%}}$, on the product of the production cross section and branching fraction ($\sigma \mathcal{B}$) for $\PH(\PSA) \to \PZ\PSA(\PH) \to \ell\ell\bbbar$, evaluated using the \CLs criterion~\cite{Junk:1999kv,LEP-CLs} in the asymptotic approximation~\cite{Cowan:2010js} as a function of the \PH and \PSA and mass hypotheses.
Model dependent exclusion regions at 95\% \CL in the $\mH$~vs.~$\mA$ plane can be obtained by comparing ${\sigma_{95\%}}$ to the theoretical cross section predicted by a particular model. Figure~\ref{fig:limits_2D} shows the expected and observed 95\% \CL exclusion regions for the Type-II 2HDM benchmark scenario ${\tan\beta=1.5}$ and ${\cos (\beta - \alpha)=0.01}$, while Fig.~\ref{fig:limits_2D_tb_cba} shows the 95\% \CL exclusion region in the $\tan\beta$~vs.~${\cos (\beta - \alpha)}$ plane for ${\mH=379}$\GeV and ${\mA=172}$\GeV.

\clearpage

\section{Summary}

This paper reports on a search for a new CP-even (odd) neutral Higgs boson, decaying
into a \PZ boson and a lighter CP-odd (even) neutral Higgs boson, where the \PZ decays into an electron or muon pair, and the light Higgs boson into a {\cPqb} quark pair. The search is based on
LHC proton-proton collision data at a center-of-mass energy $\sqrt{s}=13$\TeV collected by the
CMS experiment during 2016, corresponding to an integrated luminosity of
35.9\fbinv.
We consider decays such as $\PH \to \PZ\PSA \to \ell\ell\bbbar$, where \PH and \PSA are the additional CP-even and -odd Higgs bosons above-mentioned, respectively, in the context of the two-Higgs-doublet model (2HDM). They are searched for in the mass range from 120 to 1000 GeV for \PH and 30 to 1000 GeV for \PSA. The search is subsequently extended to the $\PSA \to \PZ\PH \to \ell\ell\bbbar$ process via interchanging the two mass parameters.

No significant deviations from the standard model expectations are observed. Model independent upper limits on the product of cross section and branching fraction are set.
Limits are also set on the parameters of the 2HDM, assuming the Type-II formulation.
Under the specific benchmark scenario corresponding to $\tan\beta=1.5$ and $\cos (\beta - \alpha)=0.01$, regions with \mH in the range 150--700\GeV and \mA in the range 30--295\GeV with $\mH > \mA$, or alternatively for \mH in the range 30--280\GeV and \mA in the range 150--700\GeV with $\mH < \mA$ are excluded at 95\% confidence level.
Results are also interpreted in the scenario where $\mH=379$\GeV and $\mA=172$\GeV. In this context, the region with $\cos (\beta - \alpha)$ in the range $-$0.9--0.3 and $\tan\beta$ in the range 0.5--7.0 is excluded at 95\% confidence level.
With respect to previous searches, a larger region of the Type-II 2HDM parameter space is excluded.

\begin{acknowledgments}
We congratulate our colleagues in the CERN accelerator departments for the excellent performance of the LHC and thank the technical and administrative staffs at CERN and at other CMS institutes for their contributions to the success of the CMS effort. In addition, we gratefully acknowledge the computing centres and personnel of the Worldwide LHC Computing Grid for delivering so effectively the computing infrastructure essential to our analyses. Finally, we acknowledge the enduring support for the construction and operation of the LHC and the CMS detector provided by the following funding agencies: BMBWF and FWF (Austria); FNRS and FWO (Belgium); CNPq, CAPES, FAPERJ, FAPERGS, and FAPESP (Brazil); MES (Bulgaria); CERN; CAS, MoST, and NSFC (China); COLCIENCIAS (Colombia); MSES and CSF (Croatia); RPF (Cyprus); SENESCYT (Ecuador); MoER, ERC IUT, PUT and ERDF (Estonia); Academy of Finland, MEC, and HIP (Finland); CEA and CNRS/IN2P3 (France); BMBF, DFG, and HGF (Germany); GSRT (Greece); NKFIA (Hungary); DAE and DST (India); IPM (Iran); SFI (Ireland); INFN (Italy); MSIP and NRF (Republic of Korea); MES (Latvia); LAS (Lithuania); MOE and UM (Malaysia); BUAP, CINVESTAV, CONACYT, LNS, SEP, and UASLP-FAI (Mexico); MOS (Montenegro); MBIE (New Zealand); PAEC (Pakistan); MSHE and NSC (Poland); FCT (Portugal); JINR (Dubna); MON, RosAtom, RAS, RFBR, and NRC KI (Russia); MESTD (Serbia); SEIDI, CPAN, PCTI, and FEDER (Spain); MOSTR (Sri Lanka); Swiss Funding Agencies (Switzerland); MST (Taipei); ThEPCenter, IPST, STAR, and NSTDA (Thailand); TUBITAK and TAEK (Turkey); NASU (Ukraine); STFC (United Kingdom); DOE and NSF (USA).

\hyphenation{Rachada-pisek} Individuals have received support from the Marie-Curie programme and the European Research Council and Horizon 2020 Grant, contract Nos.\ 675440, 752730, and 765710 (European Union); the Leventis Foundation; the A.P.\ Sloan Foundation; the Alexander von Humboldt Foundation; the Belgian Federal Science Policy Office; the Fonds pour la Formation \`a la Recherche dans l'Industrie et dans l'Agriculture (FRIA-Belgium); the Agentschap voor Innovatie door Wetenschap en Technologie (IWT-Belgium); the F.R.S.-FNRS and FWO (Belgium) under the ``Excellence of Science -- EOS" -- be.h project n.\ 30820817; the Beijing Municipal Science \& Technology Commission, No. Z181100004218003; the Ministry of Education, Youth and Sports (MEYS) of the Czech Republic; the Lend\"ulet (``Momentum") Programme and the J\'anos Bolyai Research Scholarship of the Hungarian Academy of Sciences, the New National Excellence Program \'UNKP, the NKFIA research grants 123842, 123959, 124845, 124850, 125105, 128713, 128786, and 129058 (Hungary); the Council of Science and Industrial Research, India; the HOMING PLUS programme of the Foundation for Polish Science, cofinanced from European Union, Regional Development Fund, the Mobility Plus programme of the Ministry of Science and Higher Education, the National Science Center (Poland), contracts Harmonia 2014/14/M/ST2/00428, Opus 2014/13/B/ST2/02543, 2014/15/B/ST2/03998, and 2015/19/B/ST2/02861, Sonata-bis 2012/07/E/ST2/01406; the National Priorities Research Program by Qatar National Research Fund; the Ministry of Science and Education, grant no. 3.2989.2017 (Russia); the Programa Estatal de Fomento de la Investigaci{\'o}n Cient{\'i}fica y T{\'e}cnica de Excelencia Mar\'{\i}a de Maeztu, grant MDM-2015-0509 and the Programa Severo Ochoa del Principado de Asturias; the Thalis and Aristeia programmes cofinanced by EU-ESF and the Greek NSRF; the Rachadapisek Sompot Fund for Postdoctoral Fellowship, Chulalongkorn University and the Chulalongkorn Academic into Its 2nd Century Project Advancement Project (Thailand); the Nvidia Corporation; the Welch Foundation, contract C-1845; and the Weston Havens Foundation (USA). \end{acknowledgments}

\bibliography{auto_generated}

\providecommand{\href}[2]{#2}\begingroup\raggedright\begin{thebibliography}{10}%
\makeatletter
\providecommand{\hrefCMSnoop }[0]{\@secondoftwo}%
\makeatother
\providecommand{\doi}{\texttt{doi:}\begingroup \urlstyle{tt}\Url}

\bibitem{HiggsAtlas}
\hrefCMSnoop {}{{ATLAS Collaboration}, ``Observation of a new particle in the
  search for the {S}tandard {M}odel {H}iggs boson with the {ATLAS} detector at
  the {LHC}'',} \textit{ Phys. Lett. B} \textbf{ 716} (2012) 1,
  \href{http://dx.doi.org/10.1016/j.physletb.2012.08.020}{\doi{10.1016/j.physletb.2012.08.020}},
\href{http://www.arXiv.org/abs/1207.7214}{\texttt{arXiv:1207.7214}}.

\bibitem{HiggsCms}
\hrefCMSnoop {}{{CMS Collaboration}, ``Observation of a new boson at a mass of
  125~{GeV} with the {CMS} experiment at the {LHC}'',} \textit{ Phys. Lett. B}
  \textbf{ 716} (2012) 30,
  \href{http://dx.doi.org/10.1016/j.physletb.2012.08.021}{\doi{10.1016/j.physletb.2012.08.021}},
\href{http://www.arXiv.org/abs/1207.7235}{\texttt{arXiv:1207.7235}}.

\bibitem{HiggsCms2}
\hrefCMSnoop {}{{CMS Collaboration}, ``Observation of a new boson with mass
  near 125~{GeV} in {$\Pp\Pp$} collisions at {$\sqrt{s}$} = 7 and 8~{TeV}'',}
  \textit{ JHEP} \textbf{ 06} (2013) 081,
  \href{http://dx.doi.org/10.1007/JHEP06(2013)081}{\doi{10.1007/JHEP06(2013)081}},
\href{http://www.arXiv.org/abs/1303.4571}{\texttt{arXiv:1303.4571}}.

\bibitem{Aad:2015zhl}
\hrefCMSnoop {}{{ATLAS and CMS Collaborations}, ``Combined measurement of the
  {Higgs} boson mass in {$\Pp\Pp$} collisions at $\sqrt{s}=7$ and 8 {TeV} with
  the {ATLAS} and {CMS} experiments'',} \textit{ Phys. Rev. Lett.} \textbf{
  114} (2015) 191803,
  \href{http://dx.doi.org/10.1103/PhysRevLett.114.191803}{\doi{10.1103/PhysRevLett.114.191803}},
\href{http://www.arXiv.org/abs/1503.07589}{\texttt{arXiv:1503.07589}}.

\bibitem{Sirunyan:2017exp}
\hrefCMSnoop {}{{CMS Collaboration}, ``{Measurements of properties of the Higgs
  boson decaying into the four-lepton final state in pp collisions at
  $\sqrt{s}=13 $ TeV}'',} \textit{ JHEP} \textbf{ 11} (2017) 047,
  \href{http://dx.doi.org/10.1007/JHEP11(2017)047}{\doi{10.1007/JHEP11(2017)047}},
\href{http://www.arXiv.org/abs/1706.09936}{\texttt{arXiv:1706.09936}}.

\bibitem{Aaboud:2018wps}
\hrefCMSnoop {}{{ATLAS Collaboration}, ``{Measurement of the Higgs boson mass
  in the $\mathrm{H}\rightarrow \mathrm{ZZ}^* \rightarrow 4\ell$ and
  $\mathrm{H} \rightarrow \gamma\gamma$ channels with $\sqrt{s}=13$ TeV pp
  collisions using the ATLAS detector}'',} \textit{ Phys. Lett. B} \textbf{
  784} (2018) 345,
  \href{http://dx.doi.org/10.1016/j.physletb.2018.07.050}{\doi{10.1016/j.physletb.2018.07.050}},
\href{http://www.arXiv.org/abs/1806.00242}{\texttt{arXiv:1806.00242}}.

\bibitem{Branco:2011iw}
G.~C. Branco\hrefCMSnoop {}{ {et~al.}, ``{Theory and phenomenology of
  two-Higgs-doublet models}'',} \textit{ Phys. Rept.} \textbf{ 516} (2012) 1,
  \href{http://dx.doi.org/10.1016/j.physrep.2012.02.002}{\doi{10.1016/j.physrep.2012.02.002}},
\href{http://www.arXiv.org/abs/1106.0034}{\texttt{arXiv:1106.0034}}.

\bibitem{Csaki:1996ks}
\hrefCMSnoop {}{C.~Cs\'aki, ``{The minimal supersymmetric standard model}'',}
  \textit{ Mod. Phys. Lett. A} \textbf{ 11} (1996) 599,
  \href{http://dx.doi.org/10.1142/S021773239600062X}{\doi{10.1142/S021773239600062X}},
\href{http://www.arXiv.org/abs/hep-ph/9606414}{\texttt{arXiv:hep-ph/9606414}}.

\bibitem{Gerard:2007kn}
\hrefCMSnoop {}{J.~M. G\'erard and M.~Herquet, ``{A twisted custodial symmetry
  in the two-Higgs-doublet model}'',} \textit{ Phys. Rev. Lett.} \textbf{ 98}
  (2007) 251802,
  \href{http://dx.doi.org/10.1103/PhysRevLett.98.251802}{\doi{10.1103/PhysRevLett.98.251802}},
\href{http://www.arXiv.org/abs/hep-ph/0703051}{\texttt{arXiv:hep-ph/0703051}}.

\bibitem{Aaboud:2018eoy}
\hrefCMSnoop {}{{ATLAS Collaboration}, ``{Search for a heavy Higgs boson
  decaying into a $Z$ boson and another heavy Higgs boson in the $\ell\ell$bb
  final state in pp collisions at $\sqrt{s}=13$ TeV with the ATLAS
  detector}'',} \textit{ Phys. Lett. B} \textbf{ 783} (2018) 392,
  \href{http://dx.doi.org/10.1016/j.physletb.2018.07.006}{\doi{10.1016/j.physletb.2018.07.006}},
\href{http://www.arXiv.org/abs/1804.01126}{\texttt{arXiv:1804.01126}}.

\bibitem{Khachatryan:2016are}
\hrefCMSnoop {}{{CMS Collaboration}, ``{Search for neutral resonances decaying
  into a Z boson and a pair of b jets or $\tau$ leptons}'',} \textit{ Phys.
  Lett. B} \textbf{ 759} (2016) 369,
  \href{http://dx.doi.org/10.1016/j.physletb.2016.05.087}{\doi{10.1016/j.physletb.2016.05.087}},
\href{http://www.arXiv.org/abs/1603.02991}{\texttt{arXiv:1603.02991}}.

\bibitem{Aad:2015wra}
\hrefCMSnoop {}{{ATLAS Collaboration}, ``{Search for a CP-odd Higgs boson
  decaying to Zh in pp collisions at $\sqrt{s} = 8$ TeV with the ATLAS
  detector}'',} \textit{ Phys. Lett. B} \textbf{ 744} (2015) 163,
  \href{http://dx.doi.org/10.1016/j.physletb.2015.03.054}{\doi{10.1016/j.physletb.2015.03.054}},
\href{http://www.arXiv.org/abs/1502.04478}{\texttt{arXiv:1502.04478}}.

\bibitem{Aaboud:2017cxo}
\hrefCMSnoop {}{{ATLAS Collaboration}, ``{Search for heavy resonances decaying
  into a \PW or \PZ boson and a Higgs boson in final states with leptons and
  {\cPqb}-jets in 36\fbinv of $\sqrt{s} = 13$~TeV $pp$ collisions with the
  ATLAS detector}'',} \textit{ JHEP} \textbf{ 03} (2018) 174,
  \href{http://dx.doi.org/10.1007/JHEP03(2018)174}{\doi{10.1007/JHEP03(2018)174}},
  \href{http://www.arXiv.org/abs/1712.06518}{\texttt{arXiv:1712.06518}}.
[Erratum: \DOI{10.1007/JHEP11(2018)051}].

\bibitem{Khachatryan:2015lba}
\hrefCMSnoop {}{{CMS Collaboration}, ``{Search for a pseudoscalar boson
  decaying into a Z boson and the 125 GeV Higgs boson in $\ell^{+}\ell^{-}
  \mathrm{b}\overline{\mathrm{b}}$ final states}'',} \textit{ Phys. Lett. B}
  \textbf{ 748} (2015) 221,
  \href{http://dx.doi.org/10.1016/j.physletb.2015.07.010}{\doi{10.1016/j.physletb.2015.07.010}},
\href{http://www.arXiv.org/abs/1504.04710}{\texttt{arXiv:1504.04710}}.

\bibitem{Sirunyan:2019xls}
\hrefCMSnoop {}{{CMS Collaboration}, ``{Search for a heavy pseudoscalar boson
  decaying to a Z and a Higgs boson at $\sqrt{s} =$ 13 TeV}'',} \textit{ Eur.
  Phys. J. C} \textbf{ 79} (2019) 564,
  \href{http://dx.doi.org/10.1140/epjc/s10052-019-7058-z}{\doi{10.1140/epjc/s10052-019-7058-z}},
\href{http://www.arXiv.org/abs/1903.00941}{\texttt{arXiv:1903.00941}}.

\bibitem{Khachatryan:2016bia}
\hrefCMSnoop {}{{CMS Collaboration}, ``{The CMS trigger system}'',} \textit{
  JINST} \textbf{ 12} (2017) P01020,
  \href{http://dx.doi.org/10.1088/1748-0221/12/01/P01020}{\doi{10.1088/1748-0221/12/01/P01020}},
\href{http://www.arXiv.org/abs/1609.02366}{\texttt{arXiv:1609.02366}}.

\bibitem{cms}
\hrefCMSnoop {}{{CMS Collaboration}, ``The {CMS} experiment at the {CERN}
  {LHC}'',} \textit{ JINST} \textbf{ 03} (2008) S08004,
\href{http://dx.doi.org/10.1088/1748-0221/3/08/S08004}{\doi{10.1088/1748-0221/3/08/S08004}}.

\bibitem{Alwall:2014hca}
J.~Alwall\hrefCMSnoop {}{ {et~al.}, ``The automated computation of tree-level
  and next-to-leading order differential cross sections, and their matching to
  parton shower simulations'',} \textit{ JHEP} \textbf{ 07} (2014) 079,
  \href{http://dx.doi.org/10.1007/JHEP07(2014)079}{\doi{10.1007/JHEP07(2014)079}},
\href{http://www.arXiv.org/abs/1405.0301}{\texttt{arXiv:1405.0301}}.

\bibitem{Frederix:2012ps}
\hrefCMSnoop {}{R.~Frederix and S.~Frixione, ``Merging meets matching in
  {MC@NLO}'',} \textit{ JHEP} \textbf{ 12} (2012) 061,
  \href{http://dx.doi.org/10.1007/JHEP12(2012)061}{\doi{10.1007/JHEP12(2012)061}},
\href{http://www.arXiv.org/abs/1209.6215}{\texttt{arXiv:1209.6215}}.

\bibitem{Artoisenet:2012st}
\hrefCMSnoop {}{P.~Artoisenet, R.~Frederix, O.~Mattelaer, and R.~Rietkerk,
  ``Automatic spin-entangled decays of heavy resonances in {M}onte {C}arlo
  simulations'',} \textit{ JHEP} \textbf{ 03} (2013) 015,
  \href{http://dx.doi.org/10.1007/JHEP03(2013)015}{\doi{10.1007/JHEP03(2013)015}},
\href{http://www.arXiv.org/abs/1212.3460}{\texttt{arXiv:1212.3460}}.

\bibitem{Nason:2004rx}
\hrefCMSnoop {}{P.~Nason, ``A new method for combining {NLO} {QCD} with shower
  {M}onte {C}arlo algorithms'',} \textit{ JHEP} \textbf{ 11} (2004) 040,
  \href{http://dx.doi.org/10.1088/1126-6708/2004/11/040}{\doi{10.1088/1126-6708/2004/11/040}},
\href{http://www.arXiv.org/abs/hep-ph/0409146}{\texttt{arXiv:hep-ph/0409146}}.

\bibitem{Frixione:2007vw}
\hrefCMSnoop {}{S.~Frixione, P.~Nason, and C.~Oleari, ``Matching {NLO} {QCD}
  computations with parton shower simulations: the {POWHEG} method'',} \textit{
  JHEP} \textbf{ 11} (2007) 070,
  \href{http://dx.doi.org/10.1088/1126-6708/2007/11/070}{\doi{10.1088/1126-6708/2007/11/070}},
\href{http://www.arXiv.org/abs/0709.2092}{\texttt{arXiv:0709.2092}}.

\bibitem{Alioli:2010xd}
\hrefCMSnoop {}{S.~Alioli, P.~Nason, C.~Oleari, and E.~Re, ``A general
  framework for implementing {NLO} calculations in shower {M}onte {C}arlo
  programs: the {POWHEG} {BOX}'',} \textit{ JHEP} \textbf{ 06} (2010) 043,
  \href{http://dx.doi.org/10.1007/JHEP06(2010)043}{\doi{10.1007/JHEP06(2010)043}},
\href{http://www.arXiv.org/abs/1002.2581}{\texttt{arXiv:1002.2581}}.

\bibitem{Powheg_tt}
\hrefCMSnoop {}{S.~Alioli, S.-O. Moch, and P.~Uwer, ``{Hadronic top-quark
  pair-production with one jet and parton showering}'',} \textit{ JHEP}
  \textbf{ 01} (2012) 137,
  \href{http://dx.doi.org/10.1007/JHEP01(2012)137}{\doi{10.1007/JHEP01(2012)137}},
\href{http://www.arXiv.org/abs/1110.5251}{\texttt{arXiv:1110.5251}}.

\bibitem{Powheg_st}
\hrefCMSnoop {}{S.~Alioli, P.~Nason, C.~Oleari, and E.~Re, ``{NLO} single-top
  production matched with shower in {POWHEG}: $s$- and $t$-channel
  contributions'',} \textit{ JHEP} \textbf{ 09} (2009) 111,
  \href{http://dx.doi.org/10.1088/1126-6708/2009/09/111}{\doi{10.1088/1126-6708/2009/09/111}},
  \href{http://www.arXiv.org/abs/0907.4076}{\texttt{arXiv:0907.4076}}.
[Erratum: \DOI{10.1007/JHEP02(2010)011}].

\bibitem{Sjostrand:2014zea}
T.~Sj{\"o}strand\hrefCMSnoop {}{ {et~al.}, ``{An introduction to PYTHIA
  8.2}'',} \textit{ Comput. Phys. Commun.} \textbf{ 191} (2015) 159,
  \href{http://dx.doi.org/10.1016/j.cpc.2015.01.024}{\doi{10.1016/j.cpc.2015.01.024}},
\href{http://www.arXiv.org/abs/1410.3012}{\texttt{arXiv:1410.3012}}.

\bibitem{Khachatryan:2110213}
\hrefCMSnoop {}{{CMS Collaboration}, ``Event generator tunes obtained from
  underlying event and multiparton scattering measurements'',} \textit{ Eur.
  Phys. J. C} \textbf{ 76} (2015) 155,
  \href{http://dx.doi.org/10.1140/epjc/s10052-016-3988-x}{\doi{10.1140/epjc/s10052-016-3988-x}},
\href{http://www.arXiv.org/abs/1512.00815}{\texttt{arXiv:1512.00815}}.

\bibitem{Skands:2014pea}
\hrefCMSnoop {}{P.~Skands, S.~Carrazza, and J.~Rojo, ``{Tuning PYTHIA 8.1: the
  Monash 2013 tune}'',} \textit{ Eur. Phys. J. C} \textbf{ 74} (2014) 3024,
  \href{http://dx.doi.org/10.1140/epjc/s10052-014-3024-y}{\doi{10.1140/epjc/s10052-014-3024-y}},
\href{http://www.arXiv.org/abs/1404.5630}{\texttt{arXiv:1404.5630}}.

\bibitem{Ball:2014uwa}
\hrefCMSnoop {}{{NNPDF} Collaboration, ``Parton distributions for the {LHC Run
  II}'',} \textit{ JHEP} \textbf{ 04} (2015) 040,
  \href{http://dx.doi.org/10.1007/JHEP04(2015)040}{\doi{10.1007/JHEP04(2015)040}},
\href{http://www.arXiv.org/abs/1410.8849}{\texttt{arXiv:1410.8849}}.

\bibitem{Eriksson:2009ws}
\hrefCMSnoop {}{D.~Eriksson, J.~Rathsman, and O.~St{\aa}l, ``{2HDMC}:
  two-{Higgs}-doublet model calculator'',} \textit{ Comput. Phys. Commun.}
  \textbf{ 181} (2010) 189,
  \href{http://dx.doi.org/10.1016/j.cpc.2009.09.011}{\doi{10.1016/j.cpc.2009.09.011}},
\href{http://www.arXiv.org/abs/0902.0851}{\texttt{arXiv:0902.0851}}.

\bibitem{Agostinelli2003250}
\hrefCMSnoop {}{{GEANT4} Collaboration, ``{\GEANTfour}---a simulation
  toolkit'',} \textit{ Nucl. Instrum. Meth. A} \textbf{ 506} (2003) 250,
\href{http://dx.doi.org/10.1016/S0168-9002(03)01368-8}{\doi{10.1016/S0168-9002(03)01368-8}}.

\bibitem{Czakon:2011xx}
\hrefCMSnoop {}{M.~Czakon and A.~Mitov, ``Top++: {A} program for the
  calculation of the top-pair cross-section at hadron colliders'',} \textit{
  Comput. Phys. Commun.} \textbf{ 185} (2014) 2930,
  \href{http://dx.doi.org/10.1016/j.cpc.2014.06.021}{\doi{10.1016/j.cpc.2014.06.021}},
\href{http://www.arXiv.org/abs/1112.5675}{\texttt{arXiv:1112.5675}}.

\bibitem{Li:2012wna}
\hrefCMSnoop {}{Y.~Li and F.~Petriello, ``Combining {QCD} and electroweak
  corrections to dilepton production in the framework of the {FEWZ} simulation
  code'',} \textit{ Phys. Rev. D} \textbf{ 86} (2012) 094034,
  \href{http://dx.doi.org/10.1103/PhysRevD.86.094034}{\doi{10.1103/PhysRevD.86.094034}},
\href{http://www.arXiv.org/abs/1208.5967}{\texttt{arXiv:1208.5967}}.

\bibitem{Butterworth:2015oua}
\hrefCMSnoop {}{J.~Butterworth {et~al.}, ``{PDF4LHC recommendations for LHC Run
  II}'',} \textit{ J. Phys. G} \textbf{ 43} (2016) 023001,
  \href{http://dx.doi.org/10.1088/0954-3899/43/2/023001}{\doi{10.1088/0954-3899/43/2/023001}},
\href{http://www.arXiv.org/abs/1510.03865}{\texttt{arXiv:1510.03865}}.

\bibitem{Gehrmann:2014fva}
T.~Gehrmann\hrefCMSnoop {}{ {et~al.}, ``{$\PW^+\PW^-$} production at hadron
  colliders in next-to-next-to-leading order {QCD}'',} \textit{ Phys. Rev.
  Lett.} \textbf{ 113} (2014) 212001,
  \href{http://dx.doi.org/10.1103/PhysRevLett.113.212001}{\doi{10.1103/PhysRevLett.113.212001}},
\href{http://www.arXiv.org/abs/1408.5243}{\texttt{arXiv:1408.5243}}.

\bibitem{Campbell:2011bn}
\hrefCMSnoop {}{J.~M. Campbell, R.~K. Ellis, and C.~Williams, ``Vector boson
  pair production at the {LHC}'',} \textit{ JHEP} \textbf{ 07} (2011) 018,
  \href{http://dx.doi.org/10.1007/JHEP07(2011)018}{\doi{10.1007/JHEP07(2011)018}},
\href{http://www.arXiv.org/abs/1105.0020}{\texttt{arXiv:1105.0020}}.

\bibitem{deFlorian:2016spz}
\hrefCMSnoop {}{{LHC Higgs Cross Section Working Group}, ``Handbook of {LHC}
  {H}iggs cross sections: 4. {D}eciphering the nature of the {H}iggs sector'',}
  \textit{ CERN} (2016)
  \href{http://dx.doi.org/10.23731/CYRM-2017-002}{\doi{10.23731/CYRM-2017-002}},
\href{http://www.arXiv.org/abs/1610.07922}{\texttt{arXiv:1610.07922}}.

\bibitem{Chatrchyan2011}
\hrefCMSnoop {}{{CMS Collaboration}, ``Measurement of the inclusive {W} and {Z}
  production cross sections in pp collisions at {$\sqrt{s}=7\TeV$} with the
  {CMS} experiment'',} \textit{ JHEP} \textbf{ 10} (2011) 132,
  \href{http://dx.doi.org/10.1007/JHEP10(2011)132}{\doi{10.1007/JHEP10(2011)132}},
\href{http://www.arXiv.org/abs/1107.4789}{\texttt{arXiv:1107.4789}}.

\bibitem{Sirunyan:2017ulk}
\hrefCMSnoop {}{{CMS Collaboration}, ``Particle-flow reconstruction and global
  event description with the {CMS} detector'',} \textit{ JINST} \textbf{ 12}
  (2017) P10003,
  \href{http://dx.doi.org/10.1088/1748-0221/12/10/P10003}{\doi{10.1088/1748-0221/12/10/P10003}},
\href{http://www.arXiv.org/abs/1706.04965}{\texttt{arXiv:1706.04965}}.

\bibitem{Cacciari:2008gp}
\hrefCMSnoop {}{M.~Cacciari, G.~P. Salam, and G.~Soyez, ``The anti-\kt jet
  clustering algorithm'',} \textit{ JHEP} \textbf{ 04} (2008) 063,
  \href{http://dx.doi.org/10.1088/1126-6708/2008/04/063}{\doi{10.1088/1126-6708/2008/04/063}},
\href{http://www.arXiv.org/abs/0802.1189}{\texttt{arXiv:0802.1189}}.

\bibitem{Cacciari:2011ma}
\hrefCMSnoop {}{M.~Cacciari, G.~P. Salam, and G.~Soyez, ``{F}ast{J}et user
  manual'',} \textit{ Eur. Phys. J. C} \textbf{ 72} (2012) 1896,
  \href{http://dx.doi.org/10.1140/epjc/s10052-012-1896-2}{\doi{10.1140/epjc/s10052-012-1896-2}},
\href{http://www.arXiv.org/abs/1111.6097}{\texttt{arXiv:1111.6097}}.

\bibitem{Khachatryan:2015hwa}
\hrefCMSnoop {}{{CMS Collaboration}, ``Performance of electron reconstruction
  and selection with the {CMS} detector in proton-proton collisions at
  $\sqrt{s}$ = 8~{TeV}'',} \textit{ JINST} \textbf{ 10} (2015) P06005,
  \href{http://dx.doi.org/10.1088/1748-0221/10/06/P06005}{\doi{10.1088/1748-0221/10/06/P06005}},
\href{http://www.arXiv.org/abs/1502.02701}{\texttt{arXiv:1502.02701}}.

\bibitem{Sirunyan:2018fpa}
\hrefCMSnoop {}{{CMS Collaboration}, ``{Performance of the CMS muon detector
  and muon reconstruction with proton-proton collisions at $\sqrt{s}=$ 13
  TeV}'',} \textit{ JINST} \textbf{ 13} (2018) P06015,
  \href{http://dx.doi.org/10.1088/1748-0221/13/06/P06015}{\doi{10.1088/1748-0221/13/06/P06015}},
\href{http://www.arXiv.org/abs/1804.04528}{\texttt{arXiv:1804.04528}}.

\bibitem{Bodek:2012id}
A.~Bodek\hrefCMSnoop {}{ {et~al.}, ``Extracting muon momentum scale corrections
  for hadron collider experiments'',} \textit{ Eur. Phys. J. C} \textbf{ 72}
  (2012) 2194,
  \href{http://dx.doi.org/10.1140/epjc/s10052-012-2194-8}{\doi{10.1140/epjc/s10052-012-2194-8}},
\href{http://www.arXiv.org/abs/1208.3710}{\texttt{arXiv:1208.3710}}.

\bibitem{Khachatryan:2016kdb}
\hrefCMSnoop {}{{CMS Collaboration}, ``Jet energy scale and resolution in the
  {CMS} experiment in {$\Pp\Pp$} collisions at 8~{TeV}'',} \textit{ JINST}
  \textbf{ 12} (2017) P02014,
  \href{http://dx.doi.org/10.1088/1748-0221/12/02/P02014}{\doi{10.1088/1748-0221/12/02/P02014}},
\href{http://www.arXiv.org/abs/1607.03663}{\texttt{arXiv:1607.03663}}.

\bibitem{metPF}
\hrefCMSnoop {}{{CMS Collaboration}, ``Performance of the {CMS} missing
  transverse momentum reconstruction in {$\Pp\Pp$} data at $\sqrt{s}$ =
  8~{TeV}'',} \textit{ JINST} \textbf{ 10} (2015) P02006,
  \href{http://dx.doi.org/10.1088/1748-0221/10/02/P02006}{\doi{10.1088/1748-0221/10/02/P02006}},
\href{http://www.arXiv.org/abs/1411.0511}{\texttt{arXiv:1411.0511}}.

\bibitem{Sirunyan:2019kia}
\hrefCMSnoop {}{{CMS Collaboration}, ``{Performance of missing transverse
  momentum reconstruction in proton-proton collisions at $\sqrt{s} =$ 13 TeV
  using the CMS detector}'',} \textit{ JINST} \textbf{ 14} (2019) P07004,
  \href{http://dx.doi.org/10.1088/1748-0221/14/07/P07004}{\doi{10.1088/1748-0221/14/07/P07004}},
\href{http://www.arXiv.org/abs/1903.06078}{\texttt{arXiv:1903.06078}}.

\bibitem{Sirunyan:2017ezt}
\hrefCMSnoop {}{{CMS Collaboration}, ``{Identification of heavy-flavour jets
  with the CMS detector in pp collisions at 13 TeV}'',} \textit{ JINST}
  \textbf{ 13} (2018) P05011,
  \href{http://dx.doi.org/10.1088/1748-0221/13/05/P05011}{\doi{10.1088/1748-0221/13/05/P05011}},
\href{http://www.arXiv.org/abs/1712.07158}{\texttt{arXiv:1712.07158}}.

\bibitem{CMS-PAS-LUM-17-001}
\href {https://cds.cern.ch/record/2257069}{{CMS Collaboration}, ``{CMS}
  luminosity measurements for the 2016 data taking period'',} CMS Physics
  Analysis Summary CMS-PAS-LUM-17-001, 2017.

\bibitem{Sirunyan:2018nqx}
\hrefCMSnoop {}{{CMS Collaboration}, ``{Measurement of the inelastic
  proton-proton cross section at $ \sqrt{s}=13 $ TeV}'',} \textit{ JHEP}
  \textbf{ 07} (2018) 161,
  \href{http://dx.doi.org/10.1007/JHEP07(2018)161}{\doi{10.1007/JHEP07(2018)161}},
\href{http://www.arXiv.org/abs/1802.02613}{\texttt{arXiv:1802.02613}}.

\bibitem{Gross:2010qma}
\hrefCMSnoop {}{E.~Gross and O.~Vitells, ``{Trial factors for the look
  elsewhere effect in high energy physics}'',} \textit{ Eur. Phys. J. C}
  \textbf{ 70} (2010) 525,
  \href{http://dx.doi.org/10.1140/epjc/s10052-010-1470-8}{\doi{10.1140/epjc/s10052-010-1470-8}},
\href{http://www.arXiv.org/abs/1005.1891}{\texttt{arXiv:1005.1891}}.

\bibitem{Vitells:2011da}
\hrefCMSnoop {}{O.~Vitells and E.~Gross, ``{Estimating the significance of a
  signal in a multi-dimensional search}'',} \textit{ Astropart. Phys.} \textbf{
  35} (2011) 230,
  \href{http://dx.doi.org/10.1016/j.astropartphys.2011.08.005}{\doi{10.1016/j.astropartphys.2011.08.005}},
\href{http://www.arXiv.org/abs/1105.4355}{\texttt{arXiv:1105.4355}}.

\bibitem{Junk:1999kv}
\hrefCMSnoop {}{T.~Junk, ``Confidence level computation for combining searches
  with small statistics'',} \textit{ Nucl. Instrum. Meth. A} \textbf{ 434}
  (1999) 435,
  \href{http://dx.doi.org/10.1016/S0168-9002(99)00498-2}{\doi{10.1016/S0168-9002(99)00498-2}},
\href{http://www.arXiv.org/abs/hep-ex/9902006}{\texttt{arXiv:hep-ex/9902006}}.

\bibitem{LEP-CLs}
\href {HTTP://STACKS.IOP.ORG/0954-3899/28/I=10/A= "313}{A.~L. Read,
  ``Presentation of search results: the $\mathrm{CL_s}$ technique'',} \textit{
  J. Phys. G} \textbf{ 28} (2002) 2693.

\bibitem{Cowan:2010js}
\hrefCMSnoop {}{G.~Cowan, K.~Cranmer, E.~Gross, and O.~Vitells, ``Asymptotic
  formulae for likelihood-based tests of new physics'',} \textit{ Eur. Phys. J.
  C} \textbf{ 71} (2011) 1554,
  \href{http://dx.doi.org/10.1140/epjc/s10052-011-1554-0}{\doi{10.1140/epjc/s10052-011-1554-0}},
  \href{http://www.arXiv.org/abs/1007.1727}{\texttt{arXiv:1007.1727}}.
[Erratum: \DOI{10.1140/epjc/s10052-013-2501-z}].

\end{thebibliography}\endgroup
\cleardoublepage \appendix\section{The CMS Collaboration \label{app:collab}}\begin{sloppypar}\hyphenpenalty=5000\widowpenalty=500\clubpenalty=5000\vskip\cmsinstskip
\textbf{Yerevan Physics Institute, Yerevan, Armenia}\\*[0pt]
A.M.~Sirunyan$^{\textrm{\dag}}$, A.~Tumasyan
\vskip\cmsinstskip
\textbf{Institut f\"{u}r Hochenergiephysik, Wien, Austria}\\*[0pt]
W.~Adam, F.~Ambrogi, T.~Bergauer, J.~Brandstetter, M.~Dragicevic, J.~Er\"{o}, A.~Escalante~Del~Valle, M.~Flechl, R.~Fr\"{u}hwirth\cmsAuthorMark{1}, M.~Jeitler\cmsAuthorMark{1}, N.~Krammer, I.~Kr\"{a}tschmer, D.~Liko, T.~Madlener, I.~Mikulec, N.~Rad, J.~Schieck\cmsAuthorMark{1}, R.~Sch\"{o}fbeck, M.~Spanring, D.~Spitzbart, W.~Waltenberger, C.-E.~Wulz\cmsAuthorMark{1}, M.~Zarucki
% ...
\vskip\cmsinstskip
\dag: Deceased\\
1:  Also at Vienna University of Technology, Vienna, Austria\\
% ...
\end{sloppypar}
\end{document}